\documentstyle[psfig,12pt,tccite]{article}

\begin{document}

\title{\bf {A turning point analysis of the ergodic dynamics of
iterative maps}}
\author{P.Schmelcher and F.K.Diakonos\\Theoretische Chemie\\
Physikalisch-Chemisches Institut\\
Im Neuenheimer Feld 253\\69120 Heidelberg\\Federal Republic of Germany}

\date{\today}
\maketitle

\begin{abstract}
The dynamics of one dimensional iterative maps in the regime
of fully developed chaos is studied in detail.
Motivated by the observation of dynamical structures around the unstable 
fixed point we introduce the geometrical concept of a turning point which
represents a local minimum or maximum of the trajectory. 
Following we investigate the highly organized and structured distribution of
turning points. The turning point dynamics is discussed and the corresponding
turning point map which possesses an appealing asymptotic scaling property
is investigated. Strong correlations
are shown to exist for the turning point trajectories which
contain the information
of the fixed points as well as
the stability coefficients of the dynamical system.
For the more specialized case of symmetric maps which possess a symmetric
density we derive universal statistical properties of the corresponding
turning point dynamics. Using the turning point concept we finally develop
a method for the analysis of (one dimensional) time series.

\end{abstract}

\newpage

\section{Introduction}

In the past twenty years one dimensional iterative maps have attracted much attention and became
a very active field of research. An important reason for the strong interest in the dynamics generated
by iterative maps is the fact that they exhibit a variety of typical nonlinear phenomena such as 
period multiplication, intermittency and chaotic behaviour
[Grossmann and Thomae, 1977, Feigenbaum, 1978, Feigenbaum, 1979, Collet and Eckmann, 1980, Crutchfield et al,
1982, Ott, 1993, Schuster, 1994].
They provide therefore a tool for the modeling and simulation of the dynamics of more
complicated systems as they typically occur in physics, chemistry and biology. In particular the
chaotic dynamics has by now been shown to be of potential importance in many different fields
like for example fluids [Gollub and Benson, 1980], plasmas [Sagdeev et al, 1990],
circuits [Linsay, 1981] and lasers [Arecchi et al, 1982].
Many of the characterstic features observed in dissipative systems of higher dimensional phase space
have their origin in the universal behaviour of one dimensional processes.

Of particular relevance are the maps of the unit interval which possess a single smooth maximum. Due to
their noninvertibility they show a very rich behaviour when a control parameter is changed
continuously. A typical phenomenon is the bifurcation route to chaos and its universal scaling
laws which have been studied in detail in the literature
[Grossmann and Thomae, 1977, Feigenbaum, 1978,Collet and Eckmann, 1980]
and are now well understood. This bifurcation route leads to an attractor which is, 
with further changing control parameter, followed by an inverse band bifurcation and ends up
in the final state of fully developed chaos. It is the purpose of the present paper to investigate the  
dynamics of these maps for fully developed chaos, i.e. in their chaotic and ergodic state.
Common quantities describing this state are the
Ljapunov exponent $\lambda$ which yields the rate of exponential divergence for neighbouring trajectories
and the invariant density $\rho$ which is a measure for the distribution of orbits in the ergodic limit.
For further characterization one can use the non-uniformity factor which measures the deviation
of the rate of divergence from its average or the correlation function
[Gy\"orgyi and Szepfalusy, 1984].
Our investigation is aimed at a further understanding of the state of fully developed chaos
and goes beyond the above quantities [Diakonos and Schmelcher, 1997].
This will provide us valuable insights into the underlying
structure of the ergodic dynamics of maps. 
We emphasize that the turning point properties and concepts developed in the present paper
are valid for any one dimensional smooth and single humped map for which the rank two image
of the extremum merges into an unstable fixed point with a positive multiplier.
We shall call this rather general class of maps in the following SSH maps. The only exception is the
universal statistical behaviour derived in Sec.5 which holds only for doubly symmetric maps,
i.e. for symmetric maps which possess a symmetric density. 

In detail we proceed as follows. In chapter 2 we begin by illustrating some eyecatching features
of a typical one dimensional
ergodic trajectory. These features do { \it not} show up in the invariant density
but are related to the distribution of the oscillations of the trajectories.
We also introduce in chapter 2 the concept of turning points and discuss the role of turning
points in the ergodic dynamics.  A comparison of the chaotic map
with a random map with the same invariant density clearly reveals the structures embedded in 
the fully chaotic and ergodic dynamics. 
In chapter 3 the turning point map (TPM) is introduced and analyzed. It possesses an 
important scaling behaviour which develops by joining together humps of successive iterates of
the original map. Chapter 4 is devoted to a turning point analysis of higher iterates of the
map. Important dynamical information is shown to be present in these higher TPMs.
In particular higher periodic orbits, their preimages as well as local
stability coefficients are exhibited in the TPMs.
Chapter 5 contains a derivation of the universal statistical turning point properties for 
doubly symmetric maps.
In chapter 6 we use our turning point concept in order to develop a method for the
analysis of one dimensional time series.
Chapter 7 concludes with a summary.

\section{The dynamics and the distribution of turning points}

In the ergodic case of fully developed chaotic dynamics the measure of the chaotic trajectories
is equal to one. Unstable periodic orbits are dense in this chaotic 'sea' but they are of measure zero.
Let us begin our investigation by considering a typical chaotic trajectory.
For reasons of illustration we hereby choose the logistic map in the
ergodic limit $x(n+1) = 4 x(n) (1-x(n))$.
We emphasize however that our main observations and results of investigations are valid for
any SSH map.
Figure 1 shows such an ergodic trajectory for $1.5 \times 10^{4}$
iterations together with its invariant density for $10^{6}$ iterations. The invariant density of the 
logistic map is given by $\rho_{L} = \frac{1}{\pi \sqrt{x (1-x)}}$. Both its smoothness in the
open interval $]0,1[$ as well as its singular behaviour at $x = 0,1$ can clearly be seen in the density
of the trajectory given in Fig.1.

A closer look at the trajectory itself in Fig.1
gives us the impression that there exists a dynamical structure
in the neighbourhood of $x_F=\frac{3}{4}$ which is the position of the unstable fixed point ($x_{F}$
refers in the following always to the fixed point different from zero of the map). 
This structure possesses no counterpart in the corresponding smooth invariant density $\rho_{L}(x)$. 
Figure 2 shows the trajectory for fewer iterations which 
reveals that the above-mentioned structure in the neighbourhood of the fixed point might have
its origin in the nonuniform distribution of oscillations in the interval $[0,1]$:
we observe an enhanced probability for finding small amplitude oscillations in the
vicinity of the fixed points. 
Before investigating these properties in more detail let us introduce the following notions:
a turning point $x(n)$ of a trajectory (abbreviated by $TP$)
is a point which represents a local maximum or minimum of the
trajectory (see Eq.(1) below for a quantitative criterion for the turning point).
We remark that the name 'turning point' should not be confused with
the same name that often refers to the local extrema of a map $f$.
The center of oscillations (abbreviated by $CO$) are defined to be the midpoints between subsequent
local maxima and minima (and vice versa) and the abbreviation $AO$ means the amplitude of the
corresponding oscillations.

In the following we analyze the dynamical properties of the
turning points. This will lead us to a qualitative as well as quantitative
understanding of the mechanisms which create, among others, the corresponding
structures (see below) in the relevant densities. 
We begin by deriving a necessary and sufficient condition so that a point $x(n)$ of an ergodic
trajectory is a turning point. Our starting point is the geometric condition
\begin{equation}
(x(n+1)-x(n)) \times (x(n)-x(n-1))<0
\end{equation}
for a turning point $x(n)$ where $x(n+1)=f(x(n))$ and $f$ is a single humped and smooth
map. $x_{1}$ is the inverse image of the fixed point $x_{F}$ on the left branch $[0,x_m]$
of the map $f$ where $x_{m}$ is the maximum of the map $f$.
 We will show in the following that given a map $f$ and its fixed point $x_{F}$ the
condition for $x(n)$ to become a turning point is $x(n-1)>x_{1}$. Since the map $f$ is monotonically
increasing on $[0,x_{m}]$ and decreasing on $[x_{m},1]$ we have to distinguish three cases.
For $x(n-1)<x_{1}$ we obtain the inequality $x(n-1)<x(n)<x(n+1)$ and therefore $x(n)$ is not a turning point.
In contrast to this we obtain for $x(n-1) \in [x_1,x_F]$ the inequalities $x(n-1)<x(n)$ and $x(n)>x(n+1)$
and $x(n)$ is a turning point. The third case is $x(n-1)\in [x_F,1]$ which yields $x(n-1)>x(n)$ and
$x(n+1)>x(n)$, i.e. $x(n)$ is a turning point. In total we therefore arrive at the condition
$x(n-1)>x_{1}$ for $x(n)$ to become a turning point. Otherwise, i.e. for $x(n-1)<x_{1}$, $x(n)$ is a
non turning point. For the special case of the
logistic map this means $x(n-1)$ has to be larger than $\frac{1}{4}$ in order
for $x(n)$ to become a turning point.

With this result we are now able to understand and discuss the occurence of turning points,
i.e. we gain insights into the dynamics of turning points.
(see in particular also Fig.2 for an illustration of the turning point dynamics).
In the following we distinguish between
different phases of the motion created by the iterations of the map.
Successive iterations of the map which are located in the interval $[0,x_{1}]$ belong to a
stretching phase of the iterative motion. Apart from the very first point (see below) in the
interval $[0,x_{1}]$ the points $x(n)$ of this phase obey $x(n-1)<x_{1}$ and are therefore
non turning points.
Once we leave the interval $[0,x_{1}]$ the first point $x>x_{1}$ obeys $x\in [x_{1},x_{F}]$
and is also a non turning point. The next series of iterated points obey alternately
$x>x_F>x_1$ and $x_1<x<x_F$ and we therefore obtain an oscillating stretching phase of
motion around the fixed point $x_{F}$ which consists exclusively of turning points.
These stretching phases end in case we obtain $x>x^{*}$ where $x^{*}$ is the inverse image
of $x_{1}$ on the right branch $[x_m,1]$ of the map $f$ (for the logistic map
we have $x^{*}=\frac{3}{4}+\frac{1}{4}(\sqrt{3}-1)$). The next iterated point
$x$ is then folded back into $[0,x_{1}]$ and represents a turning point. Subsequently
the next stretching phase of non turning points in the interval $[0,x_{1}]$ starts etc.
Both the stretching phase as well as the oscillatory phase of motion can nicely be seen in
Figure 2. 

In order to gain further insights into relevant turning point properties we have illustrated in
Fig.3 the density $\rho_{TP}$ of the turning points
for a typical chaotic trajectory of the logistic map. First of all we realize that $\rho_{TP}$ exhibits
a step like structure at the position $x_F$ of the unstable fixed point. We observe an enhanced
probability of finding turning points in the interval $[x_F,1]$.
The singular behaviour of the turning point density
at $x = 0$ and $x = 1$ is, apart from a constant factor, the same
as that for the density $\rho_{L}$ and characteristic for the underlying dynamical law.

It is possible to understand the origin of the step-like structure of the
turning point density $\rho_{TP}$ in Fig.3 analytically. Using the above condition $x(n-1)>x_1$
for $x(n)$ to become a turning point we obtain
the density of the preimages of the turning points: ${\cal N} \rho_{L}(x)
\Theta (x-x_{1})$. Here $\Theta$ is the step function
$\Theta (x) = \{1~~for~~x>0;~0~~for~~x<0\}$
and ${\cal N}$ the corresponding normalization constant. Mapping this density forward with
$f$ we get the expression $\frac{\cal N}{2} \rho_{L}(x) (1+ \Theta (x-x_{F}))$ for the
density of the turning points. This short derivation clearly shows that the step like
structure of the turning point density at the position of the fixed point is a
common property for all SSH maps. Looking at the turning point density therefore
immediately reveals the positions of the fixed points different from zero for
a given SSH map.

In order to get an impression of the
dynamical information contained in Fig.3 we have to compare the turning point density
$\rho_{TP}$ of the logistic map with the density of turning points $\rho_{RTP}$
which is generated by the weighted random map.
(The weighted random map is chosen to possess the same density as the chaotic map, i.e.
$\rho_{R}=\rho_{L}$).

Indeed the density $\rho_{RTP}$ can be derived in closed form via the following
considerations. Given the tent map $g(y)=\{2y~for~y \le \frac{1}{2};~(2-2y)~if~y\ge\frac{1}{2}\}$
it is well-known that the conjugacy between $g$ and the logistic map $f$ is given by
the homeomorphism $h(y)=\frac{1-cos(\pi y)}{2}$, i.e. $h \circ g= f\circ h$.
The invariant density of $g$ is equal to one on the whole unit interval.
If $y(n)=y$ is a point of the interval, the probability that it is a turning point is
$y^2 + (1-y)^2$. The normalized random turning point distribution $\sigma_{RTP}$
belonging to the random map with unit density, i.e. $\sigma_R =1$,
is therefore given by $\sigma_{RTP}=\frac{3}{2}[y^2 + (1-y)^2]$.
The enhancement of the turning point density $\sigma_{RTP}$ compared to $\sigma_R$
at $0$ and $1$ are geometric boundary effects.
Since the conjugacy $h$ preserves the turning points and transports the invariant
density for $g$ to the invariant density of $f$ we can treat the application of $h$ as
a coordinate change of the unit interval. As a result the weighted random turning point density $\rho_{RTP}$
takes on the following appearance
\begin{equation}
\rho_{RTP}(x)=\frac{\sigma_{RTP}(y)}{\pi \sqrt{x(1-x)}}=\frac{6 [(arcsin \sqrt{x})^2+
(arccos \sqrt{x})^2]}{\pi^3 \sqrt{x(1-x)}}
\end{equation}
$\rho_{RTP}$ is symmetric with respect to
$\frac{1}{2}$ and smoothly monotonically decreasing and increasing in $]0,\frac{1}{2}]$
and $[\frac{1}{2},1[$, respectively. It possesses power
singularities at $x = 0$ and $x = 1$ which are apart from a constant
factor the same as for $\rho_{R}$.

To investigate this further
we have illustrated in Fig.4a the density for the centers of the oscillations $\rho_{LCO}$
and in Fig.4b the density for the amplitudes of the oscillations $\rho_{LAO}$ for the trajectory of the
logistic map shown in Figs.1/2. First of all we observe that the density $\rho_{LCO}$ is nonzero
only in a subset of the unit interval, namely for approximately $x \in [0.38,0.78]$.
The density for the centers of the oscillations shows three main peaks. The largest
peak is located at $\frac{1}{2}$. Two other major peaks are located at approximately $0.58$ and $0.78$, 
respectively. The centers of the oscillations are therefore not smoothly distributed but exhibit certain
dynamically prefered values.
In addition the density $\rho_{LCO}$ shows a number of small dips and step like structures. The density for the
amplitudes of the oscillations shown in Fig.4b exhibits a progression of
four sharp peaks which are preceded by step like
and strongly ascending broad structures. The positions of the peaks are approximately $0.56, 0.86, 0.96$ and $1.0$,
respectively, which represent the dynamically prefered amplitudes of the oscillations.
Both the distribution $\rho_{LCO}$ of the centers as well as the density $\rho_{LAO}$ for the amplitudes
of the oscillations of the trajectories contain therefore dynamical information which is in particular
responsible for the above-observed structures in the trajectory.
We emphasize that looking at both $\rho_{LCO},\rho_{LAO}$ with increasing resolution reveals an increasing
number of structures and detailed properties.
Part of these structures can be derived analytically by the following considerations.

Consider the points $x_0=1$ and $x_k=(f|_{[0,\frac{1}{2}]})^{-k} (x_F)$ for $k>0$
which are the preimages of the fixed point on the left branch of the map $f$.
If the point $x(n)=x$ is a turning point and $x \in (x_k,x_{k-1}]$ then the
next turning point is $x(n+k)=f^k(x)$. Set
\begin{equation}
\varphi_k (x)=\frac{f^k(x)+x}{2}
\end{equation}
which defines the center of an oscillation. Then
\begin{equation}
\rho_{LCO} = \sum_{k=1}^{\infty} (\varphi_k)_{*}(\rho_{TP} \chi_{(x_k,x_{k-1}]})
\end{equation}
where $\chi_A$ denotes the characteristic function of $A$ (1 on $A$ and 0 outside $A$).
The Perron-Frobenius operator $(\varphi_k)_{*}$ transforms the densities in the usual way
\begin{equation}
((\varphi_k)_{*}(\rho))(x)= \sum_{\varphi_k(y)=x}\frac{\rho(y)}{|\varphi_k^{'}(y)|}
\end{equation}
In particular, the support of $\rho_{LCO}$ is the union of the supports of the
densities $(\varphi_k)_{*}(\rho_{TP} \chi_{(x_k,x_{k-1}]})$, that is the union of
the sets $(\varphi_k)((x_k,x_{k-1}])$. We have
therefore for the logistic map
\begin{equation}
\varphi_1(x)=\frac{5x -4x^2}{2}
\end{equation}
and $(x_0,x_1]=(\frac{1}{4},1]$. The critical point, i.e. maximum, of $\varphi_1$ is $\frac{5}{8}$.
In addition $\varphi_1(\frac{1}{4})=\frac{1}{2}$, $\varphi_1(\frac{5}{8})=\frac{25}{32}$,
and $\varphi_1(1)=\frac{1}{2}$. Therefore $\varphi_1((x_1,x_0])=[\frac{1}{2},\frac{25}{32}]$.

If $k>1$ and $x\in (x_k,x_{k-1}]$ then $f^{k-1}(x)\in (x_1,x_F]$ and $f^{k-1}>x$ so
\begin{equation}
\varphi_k(x) = \frac{f^k(x)+x}{2}<\frac{f^k(x)+f^{k-1}(x)}{2}=\varphi_1(f^{k-1}(x))\le\frac{25}{32}
\end{equation}
On the other hand, $f^k(x)\in [x_F,1]$, so
\begin{equation}
\varphi_k(x) = \frac{f^k(x)+x}{2}>\frac{f^k(x)}{2}\ge\frac{x_F}{2}=\frac{3}{8}
\end{equation}
Therefore the support of $\rho_{LCO}$ is contained in $[\frac{3}{8},\frac{25}{32}]$.
Moreover, for $k>1$ there are points $a_k,b_k \in (x_k,x_{k-1}]$ with $f^k(a_k)=x_F$ and
$f^k(b_k)=1$. We have
\begin{equation}
\left [ \frac{a_k+x_F}{2},\frac{b_k+1}{2}\right] \subset \varphi_k((x_k,x_{k-1}])
\end{equation}
and $\frac{(a_k+x_F)}{2} \le \frac{(x_{k-1}+x_F)}{2}, \frac{b_k+1}{2}>\frac{1}{2}$, so
\begin{equation}
\left [ \frac{x_{k-1}+x_F}{2},\frac{1}{2}\right] \subset \varphi_k((x_k,x_{k-1}])
\end{equation}
Therefore the support of $\rho_{LCO}$ is equal to $[\frac{3}{8},\frac{25}{32}]$.

Apart from the support of the quantity $\rho_{LCO}$ also its singularity structure
can be derived. $\rho_{LCO}$ possesses singularities whenever $\rho_{TP}$ has a singularity
or $\varphi_k'$ is zero.
There are two singularities of $\rho_{TP}$: at $0$ and $1$. The singularity at $1$ gets
carried by $(\varphi_1)_*$ to $\frac{1}{2}$, so $\rho_{LCO}$ has a singularity at $\frac{1}{2}$.
The singularity at $0$ gets "washed out", since as $k \rightarrow \infty$, the derivative
of $\varphi_k$ is of order $2^k$, whereas $\rho_{TP}$ on $(x_k,x_{k-1}]$ is only of
order $2^{\frac{-k}{2}}$. However, the zeros of $\varphi_k'$ produce additional
singularities of $\rho_{LCO}$ (one singularity for every $k$). In particular,
$\varphi_1'(\frac{5}{8})=0$, so we get a singularity of $\rho_{LCO}$ at $\frac{25}{32}$.
For $k=2$ we have $\varphi_2'(t)=0$ for $t \approx 0.163$ (t is a solution to
$256 t^3 -384 t^2 +160 t -17 = 0$) and we get a singularity of $\rho_{LCO}$ at
$\varphi_2(t) \approx 0.577$.
To summarize, the distribution for the center of oscillations $\rho_{LCO}$
possesses a highly nontrivial structure on a finite subset of the unit interval
and shows in particular infinitely many singularities which reflect the
fact that the infinitesimal turning point neighbourhood of the origin $0$ is mapped
to arbitrary detail to the turning points in the finite neighbourhood of the
fixed point $x_F$ (see below in the context of the turning point map).

To make the structures contained in the ergodic chaotic dynamics even clearer we have illustrated 
in Fig.5a and 5b the corresponding densities $\rho_{RCO}$ and $\rho_{RAO}$ for a weighted random map 
which possesses the same density $\rho_{R}=\rho_{L}$ as the logistic map. The density for the centers
of the oscillations $\rho_{RCO}$ of the random map exhibits a single broad peak which is centered around
$\frac{1}{2}$ and decays monotonically towards the boundaries of the unit interval.
The most probable value for the position of the center of the oscillations is therefore $\frac{1}{2}$
and the probability for values different from $\frac{1}{2}$ decreases with increasing distance of
the corresponding value from $\frac{1}{2}$. 
In comparison with the density $\rho_{LCO}$ (see Fig.4a) of the logistic map whose origin is a dynamical law
the density $\rho_{RCO}$ belonging to the random map shows a very simple and smooth structure.
Figure 5b shows the density $\rho_{RAO}$ for the amplitudes of the oscillations
for the same weighted random map. $\rho_{RAO}$ is monotonically increasing
on the unit interval with a strongly peaked maximum at $x=1$. Within the 
statistical accuracy of Fig.5b $\rho_{RAO}$ is a smooth distribution 
without any structures. This clearly reveals that the fully chaotic system contains dynamical information
according to the distributions $\rho_{LCO}$ and $\rho_{LAO}$ which is not present
in the corresponding distributions $\rho_{RCO}$ and $\rho_{RAO}$ of the random map.  

To complete our picture of the oscillatory motion of the random map on the one hand and 
of a 1D dynamical law on the other hand we have illustrated in Fig.6 the occupied coordinate space for
the amplitudes of the oscillations as a function of the center of the oscillations for both the weighted random
map $x_{RAO}(x_{RCO})$ in Fig.6a as well as the logistic map $x_{LAO}(x_{LCO})$ in Fig.6b.
For the case of the random map, $x_{RAO}(x_{RCO})$ fills the two-dimensional area of an isosceles triangle (see
Fig.6a). In contrast to this the corresponding quantity $x_{LAO}(x_{LCO})$ for the logistic map
is a one dimensional curve (see Fig.6b) which consists, apart from a cross diagonal line, of a series
of skew and increasingly peaked humps. With increasing absolute values of these humps their widths tend
to zero, i.e. the series of humps approaches an accumulation line which is the limiting line of the
upper left part of the curve shown in Fig.6b. Again it is obvious that there exists a major difference
between the 'motion' of the random map which is distributed according to its weight and the highly
organized motion of the logistic map which shows according to Fig.6b a number of interesting structures.

We emphasize that the above-observed properties and behaviour of the oscillatory motion of maps
are by no means restricted to our specific example, namely the logistic map: 
all SSH maps show a very similar behaviour. This includes 
the special family of symmetric beta maps which has been introduced recently 
[Diakonos and Schmelcher, 1996] and possesses members
with a wide range for the order of the maximum and for the singular behaviour of the densities at the
positions $x=0,1$. 

Let us now investigate how subsequent turning points are mapped unto each other.

\section{The turning point map and its scaling properties}

Having discussed the occurence of turning points in the dynamics of the map $f$ we 
introduce now the turning point map (TPM), i.e. the map of a turning point onto its
subsequent turning point. The TPM will be a key quanitity for our understanding
of the structures and organization in the ergodic dynamics of iterative maps.
Figure 7 shows the TPM belonging to the logistic map on a
logarithmic scale. For reasons of illustration we have chosen the logistic map as
a specific example. However, as already emphasized our arguments
are valid for any SSH map.
The monotonous part of the TPM for $x_F < x(n) < 1$ has its origin in the fact
that the turning points in the interval $[x_F,1]$ are mapped via the TPM onto the
interval $[0,x_F]$.
The first hump in Fig.7 represents the image of the interval $[x_{1},x_{F}]$
with respect to the TPM. 
Since turning points in the interval $[x_{1},1]$ are mapped directly onto turning points by the
map $f$ itself the TPM in $[x_{1},1]$ simply reflects the shape of $f$ in this interval.

The eyecatching feature of the TPM for $x<x_1$ is the extreme similarity of its subsequent humps
with decreasing values of $x(n)$ on a logarithmic scale. Approaching
smaller and smaller values for $x(n)$ Fig.7 suggests an asymptotically exact scaling
law for subsequent humps. The TPM possesses an accumulation point of its humps 
at the origin and has dimension 1. For values $x<x_1$ the TPM shows us how turning points in the
neighbourhood of the origin $U_{\epsilon}(0)$ (which have their preimage in the neighbourhood
$U_{\delta}(1)$ of $1$) are mapped onto the turning points in the finite interval $[x_F,1]$. 
The TPM does this according to Fig.7 to arbitrary detail and in a self repeating form if we approach
the origin. 

As indicated in Fig.7 the vertical positions on the axis $x(n+1)$
of the turning points $\{x_i\}=(x_1,x_2,x_3,x_4,...)$
of the TPM are all the same and given by the fixed point $x_F$. The horizontal positions of these
turning points on the axis labeled $x(n)$
are the positions of successive preimages of the fixed point $x_F$ on the left branch
of the map $f$. The abscissae of $x_1$ is therefore the preimage $f^{-1}(x_F)$ of the fixed point $x_F$ in the
interval $[0,x_m]$,
the abscissae of $x_2$ is the preimage $f^{-1}(x_1)$ of $x_1$ in the interval $[0,x_m]$, etc...
This series of preimages obviously converges towards the unstable fixed point $0$ and the ratio of
subsequent preimages converges towards the derivative of the map $f$ at the origin, i.e. 
we have
\begin{equation}
\kappa = f'(0) = \lim_{i \rightarrow \infty} \frac{x_{i+1}}{x_i}
\end{equation}
For the special case of the logistic map we have $x_i =\frac{1}{2}( 1 - \frac{1}{2} 
\sqrt{2 + \sqrt{2 + ....\sqrt{2+ \sqrt{3}}}})$ with $i-1$ nested roots. 

But not 'only' the turning points of the TPM $\{x_i\}$ scale asymptotically with a factor $\kappa$
but in particular also successive humps defined in the intervals $\Delta_{i} = [x_{i+1},x_{i}]$ converge
asymptotically with the same ratio against a final shape. 
A numerical as well as analytical closer investigation of the i-th hump of the TPM in
the interval $\Delta_{i}$ reveals that it is identical to the top of the first hump of the i-th iterate $f^{i}$
of the map $f$. More precisely: the i-th hump of the TPM is identical to the
intersection $x(n+1)>x_{F}$ of $f^{i}$ which is closest to the unstable fixed point $0$.
Putting everything together the TPM  is given by
\begin{equation}
f_{tp}= f^{i+1}(x) \hspace*{0.5cm} \forall x\in\Delta_{i}, \hspace*{0.4cm} \forall i\ge 0
\end{equation}
with $\Delta_{0}=[x_{1},1]$. Approaching the origin the TPM can therefore be considered as a sequence
of the top of the humps of higher and higher iterated maps. We remark that even the position of the
humps of the TPM is the same as for the corresponding iterates $f^{i}$.

In addition the series of humps of the TPM converges, apart from the scaling, against an asymptotic
final shape. This asymptotic form can be obtained by using the scaling property of the humps
$f^{n+1}(x)=f^{n}(\delta_{n}x)$ which are now defined on the interval $0 \le x \le s_{n+1}$
with $s_{n+1}$ being the first zero, i.e. the zero closest to the origin, of the iterate $f^{n}$.
$\delta_{n}=\frac{s_{n}}{s_{n+1}}$ is the ratio of the first
zeros of successive iterates of the map $f$. Introducing a new rescaled variable $z=\frac{x}{s_{n+1}}$
which is defined on the whole unit interval the renormalized asymptotic functional form of the 
complete first hump is given by
\begin{equation}
h(z) = \lim_{n \rightarrow \infty} f^{n} (z s_{n})
\end{equation}
This equation provides a method for the systematic approximation of the asymptotic form $h(z)$
by calculating successive iterates of the map as well as their first zeros. The convergence
properties of this procedure turn out to be excellent.

Having understood much of the dynamics of the turning points and in particular the structure of the 
turning point map we turn now to the question of the correlation of the dynamics of the turning points.
It is well-known [Gy\"orgyi and Szepfalusy, 1984] that the strongest decay of the
correlation function occurs for the ergodic limit of the doubly symmetric maps.
They possess $\delta$-correlated
processes, i.e. their subsequent steps are completely uncorrelated. Using the common definition
of the correlation function $C_{f}(n) = \langle (x-\langle x \rangle_{f})(f^{n}(x)-\langle x \rangle_{f})\rangle_{f}$
this means $C_{f}(n) = C_{f}(0) \delta_{n0}$. Fig.8 shows now the absolute value of the
correlation function of the dynamics of
the turning points in comparison with the corresponding correlation function of a typical ergodic trajectory
for the case of the logistic map. As expected the correlation function of the chaotic trajectory decays
completely with the first time step.
(For a non doubly symmetric map the generic decay behaviour can be much 'slower').
In contrast to this, however, the correlation function of the 
turning points shows only within the first two steps a minor decay and stays subsequently {\bf constant} !
This reveals that the turning points obey a highly correlated dynamics with infinite correlation
length thereby suggesting the interpretation of the turning points as a constituent (phase) of the dynamics
which is at a critical point of a phase transition [Beck and Schl\"ogl, 1993].

Let us briefly summarize and comment on our previous results. 
Our investigation and discussion of the turning point dynamics, the turning point map as well as the 
correlation function clearly show that a turning point analysis extracts valuable dynamical information
from a given ergodic and fully chaotic system. The TPM contains details of the iterates $f^{i}$
for all $i$ and maps in particular arbitrarily small intervals in the neighbourhood of the
origin onto the finite interval $[x_F,1]$. It obeys an asymptotic scaling law and converges to 
a repeating functional form which can be derived according to the above-given construction scheme.
The fixed points as well as its preimages $x_F$ (see in particular also chapter 4
which provides an investigation of the turning point maps for higher iterates of the map $f$)
play hereby a particularly important role.  The turning point dynamics shows an
infinite correlation length of the corresponding correlation function, i.e. the turning point
analysis extracts the correlated part from the original trajectory.
The origin of this strong correlation is the oscillating dynamics around the fixed point $x_F$. 

In extension of the construction scheme for the TPM it is possible to define so-called
renormalized limit functions which are the sequence of 
suitably scaled humps of the infinitely high iterates of the map
$f$. These limit functions include not only the first hump (see Eq.(13)) but all oscillations of the infinitely high
iterates $\displaystyle{\lim_{n \rightarrow \infty} f^{n}}$ of the original map $f$.
To achieve this we define
\begin{equation}
f^{n}_k (z) = f^{n}((r_k^{(n)} - r_{k-1}^{(n)}) z + r_{k-1}^{(n)}) \hspace{0.5cm} {\rm for}
 \hspace{0.2cm} z \in [0,1]
\end{equation}
where $r_k^{(n)}$ is the k-th zero of the corresponding iterate $f^{n}$. Equation (14) scales neighbouring
zeros which converge to the same value with increasing iterations to the unit interval.
The desired renormalized family of limit functions $g^{(k)}(z)$ which describe the humps between the zeros $r_{k-1}^{(n)}$
and $r_k^{(n)}$ for $n \rightarrow \infty$, i.e. of the infinitely high iterate of $f$, is then given by the limit process
\begin{equation}
g^{(k)}(z) = \lim_{n \rightarrow \infty} f^{n}((r_k^{(n)} - r_{k-1}^{(n)}) z + r_{k-1}^{(n)})
 \hspace{0.5cm} {\rm for} \hspace{0.2cm} z \hspace{0.2cm} \in [0,1]
\end{equation}
Let us illustrate the above procedure for the example of the logistic map in the ergodic limit.
The $n-th$ iteration $x(n)$ of a point $x(0)$ can for this case be given analytically [Pincherle, 1920, Kadanoff, 1983]
\begin{equation}
x(n) = \frac{1}{2} \left[ 1 - cos \left( 2^n arccos \left( 1 -2 x(0) \right) \right) \right]
\end{equation}
The zeros $r_k^{(n)}$ take on the following appearance
\begin{equation}
r_k^{(n)} = \frac{1}{2} \left[ 1 - cos \left( \frac{k \pi}{2^{n-1}} \right)\right]
\end{equation}
Applying the renormalization and limiting procedure according to Eq.(15) the following 
result for the limit function $g^{(k)}_L(z)$ of the logistic map can be derived
\begin{equation}
g^{(k)}_L(z) = \frac{1}{2} \left[ 1 - cos \left( 2 \pi \sqrt{\left(k-1\right)^2 + \left(2k-1\right) z} \right) \right]
\hspace*{1.0cm} {\rm for} \hspace*{0.3cm} k=1,2,... \hspace*{0.3cm} {\rm and} \hspace*{0.3cm} z \in [0,1]
\end{equation}
The next chapter is devoted to a turning point analysis of doubly iterated maps $f^{2}$ and will
give us a general idea of the information contained in arbitrarily high iterates $f^{n}$.

\section{Dynamical information of the turning point maps for higher iterates of the map}

Let us begin our investigation of the TPMs belonging to higher iterates $f^{i}$ by deriving the general 
turning point criterion. According to Eq.(1) turning points occur if
\begin{equation}
{\cal A}(x(n-1)) = (F(F(x(n-1)))-F(x(n-1))) \times (F(x(n-1))-x(n-1)) < 0
\end{equation}
where $F=f^{i}$ can now be any iterate of the original map $f$.
In order to determine the transition from turning points to non-turning points or vice versa
in the criterion Eq.(19) we have
to look for those zeros $x(n-1)$ of ${\cal A}$ for which ${\cal A}$ changes
from positive to negative values
or vice versa, i.e. the zeros with $\frac{\partial {\cal A}}{\partial x} (x(n-1)) \ne 0$.
If $x(n-1)$ is a fixed point we have both ${\cal A} (x(n-1)) =0$ and $\frac{\partial {\cal A}}{\partial x}(x(n-1)) = 0$
and $x(n-1)$ is not a point which separates turning points from non turning points with respect to $x(n)=F(x(n-1))$.
However, if $x(n-1)$ is the preimage $p_i$ of a fixed point and consequently $x(n)$ a fixed point we obtain
${\cal A} (x(n-1)) =0$ and in particular
\begin{equation}
\frac{\partial {\cal A}}{\partial x}(x(n-1)) = (x(n-1) - x(n)) \times F'(x(n-1))
\times (1 - F'(x(n)) \ne 0
\end{equation}
since $|F'(x(n))|>1$ (we have only unstable fixed points) and we consider only those preimages of the
fixed points which possess a derivative $F'(x(n-1)) \ne 0$.  This means that the
preimages of the fixed points determine the transition points from turning points to non turning points
or vice versa. We therefore obtain the following turning point criterion.
The preimages $\{p_i\}$ of the fixed points define a unique partitioning of the unit interval $[0,1]$ 
into subintervals $I_i = [p_i,p_{i+1}]$. 
In order for $x(n)$ to become a turning point $x(n-1)$ has to obey the necessary and sufficient condition 
$\displaystyle{x(n-1) \in \cup_{i=0}^{l} I_{2i+1}}$.

Let us illustrate this for the example of the twice iterated logistic 
map $F=f^{2}$. $f^{2}$ possesses four fixed points, namely $\{x_{Fi}\} = \{ 0, \frac{5-\sqrt{5}}{8},
\frac{5+\sqrt{5}}{8}, \frac{3}{4}\}$ with alternating positive and negative derivatives ${f^{2}}'(x_{Fi})$,
respectively. The preimages of the fixed points which obey the above condition $F'(p_i) \ne 0$ are
\begin{eqnarray}
p_1 &=& (\frac{1}{2}-\frac{1}{2} \sqrt{\frac{5+\sqrt{5}}{8}}) \hspace*{2.0cm}
p_2 = \frac{1}{2}(1 - \frac{\sqrt{3}}{2}) \nonumber \\
p_3 &=& \frac{1}{2} (\frac{3}{4} - \frac{\sqrt{5}}{4}) \hspace*{3.0cm}
p_4=(\frac{1}{2}-\frac{1}{2} \sqrt{\frac{5-\sqrt{5}}{8}}) \nonumber \\
p_5 &=& \frac{1}{4} \hspace*{5.0cm}
p_6 = \frac{1}{2} (\frac{3}{4} + \frac{\sqrt{5}}{4}) \nonumber \\
p_7 &=& (\frac{1}{2}+\frac{1}{2} \sqrt{\frac{5-\sqrt{5}}{8}}) \hspace*{2.0cm}
p_8=\frac{1}{2} (1 + \frac{\sqrt{3}}{2}) \nonumber \\
p_9 &=& (\frac{1}{2}+\frac{1}{2} \sqrt{\frac{5+\sqrt{5}}{8}}) \hspace*{2.0cm}
\end{eqnarray}
and in addition we have for the first and last point $p_0=0$ and $p_{10}=1$, respectively.
$x(n-1)$ has therefore to be located in one of the intervals $I_1,I_3,I_5,I_7$ or $I_9$ if we 
want $x(n)$ to become a turning point.

Before beginning with the discussion of TPM's of higher iterates $f^{n}$ some remarks are appropriate.
There exist two different classes of fixed points for our map $f^{n}$. The first class contains the fixed points with
a positive derivative, i.e. ${f^{n}}'(x_{Fi})>1$. The local dynamics around these fixed points can be characterized
as an unstable stretching behaviour. The dynamics in the neighbourhood of these points contains no turning points.
The second class consists of the fixed points with a negative derivative, i.e. ${f^{n}}'(x_{Fi})< -1$.
The local dynamics around these fixed points can be characterized as an alternating stretching behaviour which
means that successive points are on opposite sides of the corresponding fixed point. Each point of this local
dynamics is a turning point. Having these facts in mind we turn now to our discussion of the TPM of
doubly iterated maps. 

Figure 9 shows the TPM of the doubly iterated logistic map on a logarithmic scale. First of all we observe
that there are a number of disconnected objects each of which exhibits
a detailed and relatively complicated structure.
For $x(n)>y_3$ we observe the objects labeled by $(b),(c)$ and $(d)$ which appear only once.
In contrast to this we observe for $x(n)<y_3$ with decreasing values of $x(n)$ a repeating
sequence of structures. The absolute unit of these repeating structures consists of two non-overlaping
objects which are labeled in Fig.9 by $(a)$ and $(e)$, respectively. Since the repetition of these
structures on the logarithmic scale becomes asymptotically exactly periodic we obtain an asymptotic scaling
behaviour with respect to the above-mentioned absolute unit consisting of the asymptotic form of the objects
$(a)$ and $(e)$. This scaling behaviour is similar to the one observed in
the previous chapter for the TPM of the map $f$ itself. The scaling factor is the derivative of the
map $f^{2}$ at the origin namely ${f^{2}}'(0)$, i.e. the stability of the fixed point $0$.
Before entering the discussion of each of the objects $(a),(b),(c),(d)$
separately, we want to extract some important properties from the TPM of the doubly iterated map $f^{2}$ which
can easily be extended to arbitrary iterates $f^{i}$. As indicated in Fig.9 the vertical positions
of the lower and upper horizontal edges of the objects $(a),(b),(c),(d),(e)$ etc. are given by the positions of the 
fixed points $x_{F1},x_{F2},x_{F3},x_{F4}$ and the point $1$. The valuable information of the positions of the 
fixed points can therefore easily be extracted from the corresponding TPM. Higher periodic orbits can
simply be localized by extracting the positions of the fixed points from
the TPM of higher iterates of the map. In our case of Fig.9 
$x_{F1}$ and $x_{F3}$ are fixed points of $f$ whereas $\{x_{F2},x_{F4}\}$ represents a period two orbit 
of $f$.

The TPM of Fig.9 still contains much more information. Let us consider the vertical right and left edges
of the individual objects which are part of the TPM. The positions of these edges are indicated in Fig.9
by $y_1,y_2,...,y_6$. These quantities $y_i$ are the positions of the first or higher preimages
of the fixed point $x_{F3}$. $y_1,y_2,y_3,y_4$ are first preimages whereas $y_5,y_6$ are second
preimages of $x_{F3}$. In general these vertical edges are given by the positions of the preimages
of the fixed points with a positive derivative ${f^{i}}'(x_{Fi})>0$. Approaching smaller and
smaller values for $x(n)$ we obtain increasingly higher preimages of the corresponding fixed points. After having
discussed the positions of the individual objects in the TPM let us now turn to an investigation 
of the objects themselves. 

Apart from the objects of type $(e)$ all constituents of the TPM in Fig.9 consist of a series
of humps. The widths of the humps belonging to the objects of type $(a)$ decrease strongly if
the values of $x(n)$ approach the corresponding values of the preimages $y_i$ of the fixed point
$x_{F3}$. The objects $(b)$ and $(c)$ possess such an accumulation point of humps only on
one side, namely at the position of the preimage $y_2$. In order to analyze this series of
humps with decreasing widths we have illustrated in 
Figure 10 important parts of the objects labeled by $(a),(b),(c),(d)$ in Fig.9.
Figure 10 (a) shows the object $(a)$ of Fig.9 in the neighbourhood of the preimage $y_3$
of $x_{F3}$, i.e. $x(n+1)$
is illustrated as a function of $y_3 - x(n)$ with a logarithmic abscissae. 

The eyecatching feature is the asymptotically ($x(n) \rightarrow y_3$) exact 
scaling behaviour of the individual humps with a constant scaling factor.
We therefore encounter not only the above-mentioned scaling behaviour of the series of objects
of type $(a),(e)$ but also a scaling behaviour of the interior of the objects 
themselves (apart from $(e)$). The corresponding interior scaling factor is in the case
of Fig.10 (a) ${f^{2}}'(x_{F3})$, i.e. the derivative of the map $f^{2}$ at the
position of the fixed point $(x_{F3})$ with a positive derivative.
The lower turning points $\delta y_{3i}$ for $i=1,2,3$ indicated in Fig.10 (a)
represent a series of increasingly higher preimages of the fixed point $x_{F4}$ which
possesses a negative derivative ${f^{2}}'(x_{F4})<0$. This
series converges in an asymptotically scaling way towards the preimage $y_3$ of $x_{F3}$.
The structure of the object $(a)$ can be explained in even more detail. The series of humps
can be obtained by intersecting increasingly higher iterates $f^{2n}$ of the map $f^{2}$ with the
line $x(n+1)=x_{F4}$ and joining together the resulting humps. In this sense the object $(a)$
describes the top $x(n+1)>x_{F4}$ of the humps of arbitrarily high iterates of the map $f^{2}$ 
in the interval $[y_4,y_3]$. Most of the properties discussed for the example of the individual
object labeled $(a)$ can also be observed for the other constituents of the TPM in Fig.9.

Figure 10 (b) shows the object labeled $(b)$ in Fig.9 in the neighbourhood of the preimage
$y_2$, i.e. $x(n+1)$ as a function of $y_2 - x(n)$ with a logarithmic abscissae.
The same asymptotic scaling behaviour now for $x(n) \rightarrow y_2$ can be observed and 
the scaling factor is again determined by the stability of the fixed point $(x_{F3})$ which
possesses a positive derivative.
The upper turning points $\delta Y_{2i}$ for $i=1,2,3$ indicated in Fig.10 (b)
represent a series of increasingly higher preimages of the fixed point $x_{F2}$ which
possesses a negative derivative ${f^{2}}'(x_{F2})<0$. This
series converges in an asymptotically scaling way towards the preimage $y_2$ of $x_{F3}$.
The series of reversed humps can be obtained by intersecting increasingly higher iterates
of the map $f^{2}$ with the line $x(n+1)=x_{F2}$ and taking the lower part $x(n+1)<x_{F2}$.
Figures 10 (c) and (d) illustrate the structures of the objects $(c)$ and $(d)$ in Fig.9.
Asymptotic scaling properties are again observed and described by the stability of the fixed
point with a positive derivative. For Fig.10 (c) the lower turning points $\delta y_{2i}$ for $i=1,2,3$
which represent increasingly higher preimages of the fixed point $x_{F4}$ converge towards
the preimage $y_2$ of the fixed point $x_{F3}$. The humps of the object $(c)$ can be explained
analogously to the humps in Fig.10(a). Finally Fig.10 (d) shows a scaling of reversed humps
towards the preimage $y_1$ and another series of the preimages $\delta y_{1i}$ of the fixed point $x_{F3}$
which converges to $y_1$. 

Let us briefly summarize the observations. Turning point maps belonging to higher iterates of the
original map $f$ contain a number of dynamically most relevant informations. It is 
easy to extract the position of not too long periodic orbits, their stability coefficients
as well as the (higher) preimages of the fixed points of $f$ by considering the
TPM of the corresponding iterate $f^{n}$. 
The TPMs themselves as well as their disconnected constituents show an asymptotic scaling behaviour
and exactly describe important parts of arbitrarily high iterates of the map $f$.

To conclude our discussion of the TPMs of higher iterates $f^{n}$ we show in Fig.11 the 
density $\rho_{TP2}(x)$ of turning points for the doubly iterated map $f^{2}$.
We observe step like structures at the positions of the fixed points $x_{Fi}$ which reflect the
above-discussed properties of the corresponding turning point map. The singular behaviour
of $\rho_{TP2}(x)$ at the boundaries of the unit interval is apart from a constant factor
the same as for the density of the map itself.

\section{Statistical properties of the turning point dynamics}

Having studied in detail the dynamical properties of the turning points in the previous chapters
we turn now to an investigation of the statistical turning point properties.
In contrast to the previous Secs. the results of the present Sec. hold only
for doubly symmetric maps.
A measure for the frequency of the turning points is given by the ratio of the number of turning points
$N_{tp}$ and the total number of points $N_{tot}$, i.e. $P_{tp}=\frac{N_{tp}}{N_{tot}}$.
One half of that number (that is, only minima counted) has been called wave number
[Ding, 1988] or over rotation number [Blokh and Misiurewicz] in the literature.
Since we are dealing with a fully chaotic and ergodic system the quantity $P_{tp}$ is solely determined
by the necessary and sufficient condition $x(n-1)>x_{1}$ for a turning point $x(n)$ and the
frequency with which a certain region of the interval is visited. This frequency is however
given by the invariant density $\rho_{f}(x)$ of the map $f$. We therefore obtain the
following expression for the portion of turning points of the chaotic trajectories
\begin{equation}
P_{tp}=\int_{x_1}^{1} \rho_{f}(x)\, dx
\end{equation}
where $x_{1}=(1-x_{F})$ is the inverse
image of the fixed point $x_{F}>\frac{1}{2}$ of the map $f$.
The question now arises how strong the quantity $P_{tp}$ varies if we choose
different maps $f$. Since all doubly symmetric maps are smoothly conjugate to each other
we want to study the behaviour of the quantity $P_{tp}$ under conjugation. Let the conjugation
$u$ transform the map $g$ into $f$, i.e. $f= u \circ g \circ u^{-1}$. The invariant measure of $g$
is then given by $\mu_{g}(x)=\mu_{f}(u(x))$. It follows that the density transforms according
to $\rho_{g}(x)=\rho_{f}(u(x)) \frac{du}{dx}$ and we therefore obtain
\begin{equation}
P_{tp}=\int_{u^{-1}(1-x_{F})}^{1} \rho_{g}(y)\, dy
\end{equation}
The lower integration limit $u^{-1}(1-x_{F})$ can be calculated in the following way.
Since $x_F = f(x_F) = u \circ g \circ u^{-1}(x_F)$ we obtain $u^{-1}(x_F) = g (u^{-1}(x_F))$
which means that $u^{-1}(x_F)$ is a fixed point of $g$. In addition the conjugation $u$
possesses the property $u(x)=1-u(1-x)$ which is due to the fact that we are dealing with
symmetric maps. The image of the fixed point $x_{F}$ of $f$ is therefore also a fixed point of $g$,
i.e. $y_{F}=u^{-1}(x_{F})$ and we arrive at the final equation
\begin{equation}
P_{tp}=\int_{y_{1}}^{1} \rho_{g}(y)\, dy
\end{equation}
with $y_{1}=(1-y_{F})$.
This proves the invariance of the statistics of turning points under conjugation. All
doubly symmetric maps possess therefore the same ratio of turning points and non turning
points and $P_{tp}$ is therefore a universal statistical quantity.
Calculating this integral for the specific case of the logistic map yields $P_{tp}=\frac{2}{3}$
which simply means that statistically two out of three points of the trajectories are turning
points.

\section{Application of the turning point concept to the analysis of one dimensional time series}

The above-discussed characteristics of the turning point dynamics in the neighbourhood of the
unstable fixed points of the chaotic system can in fact be used to analyze chaotic time series.
The main goal of such an analysis is to find the location of the fixed points and higher periodic
orbits of the underlying dynamical system.

Since we are interested in fully chaotic and ergodic
systems we encounter only hyperbolic fixed points (HFP). We have however to distinguish between
HFP with and without reflection. For HFP with reflection subsequent iterations of the map in
the vicinity of the fixed point oscillate around the fixed point whereas for HFP without reflection
a turning point in the neighbourhood of the fixed point is always followed by
a stretching phase away from the fixed point. We will describe below an algorithm which allows us to determine
the positions of the fixed points (periodic orbits) of a chaotic time series using these properties.
The most important feature of our turning point method for the location of the fixed points is the
fact that the above-described properties {\it are not restricted to the linear neighbourhood of the
fixed point}. Therefore the position of the fixed point can be approximately determined even in the 
case when the linear neighbourhood of the fixed point is not visited by the finite trajectory, i.e.
time series, at all.

As a first step of our algorithm we select those points of a given time series $\{x(i)| i=1,...,N\}$
which are turning points, i.e. which obey Eq.(1). 
The next step distinguishes between HFP with and without reflection. To determine the HFPs with 
reflection we select those points $\{x_t(i)\}$ of the turning point trajectory (resulting from
the first step) which have as a next iteration point in the original time series also a 
turning point $\{x_t(i+1)\}$. We group these points according to the values of their center
of oscillations defined as $x_{CO}(i)=\frac{1}{2}(x_t(i)+x_t(i+1))$. Thereby points with 
approximately the same values for $x_{CO}(i)$ belong to the same group. Using the amplitudes of oscillations
$d(i)=|x_t(i+1)-x_t(i)|$ we look subsequently within each group for the turning point pair
$(\bar{x}_t(i),\bar{x}_t(i+1))$ which corresponds to the minimal amplitude of oscillations.
The position of the corresponding HFP is then given by the corresponding center of oscillation
$x_{F}=\frac{1}{2}(\bar{x}_t(i)+\bar{x}_t(i+1))$.
By using the minimal amplitude we get rid of the big oscillations which occur due to the effects of the boundaries
of the finite interval.

We emphasize that the minimal amplitude of oscillations needs not to be smaller than the size of the
linear neighbourhood of the fixed point in order to obtain a good estimation of the position 
of the fixed point. Our method possesses therefore an advantage
compared to methods relying on the properties
of the linear regime around the fixed point or recurrence methods. This is due to the fact
that the turning around the fixed point is {\it{not}} a strongly localized property, and in particular
not restricted to the linear regime. 

For the HFP without reflection we collect those points of the turning point trajectory which have as 
subsequent iteration points in the original trajectory non turning points, i.e. points belonging to 
a stretching phase of motion. If the stretching phase is increasing the value of the coordinate,
the corresponding points belong to the group $\{x_t^{+}(i)\}$ and in the opposite case to
$\{x_t^{-}(i)\}$. As a next step we look for the extrema defined by $x_F^+ = min\{x_t^{+}(i)\}$
and $x_F^- = max\{x_t^{-}(i)\}$. The position of the fixed point is finally given by
$x_F=\frac{1}{2}(x_F^+ + x_F^-)$. 

In order to demonstrate the efficiency of our algorithm we have applied it to a time series
of the second iterate of the logistic map $f^{2}(x)= f(f(x))$ with $f(x)=rx(1-x)$ for $r=3.9$.
Using a times series of only $200$ points we are able to determine the positions of the fixed points
of the system with a relative
accuracy of $0.004$. For the positions of the two HFP with reflection we get the values
$x_{F1}=0.36052$ and $x_{F2}=0.89876$ compared to the exact values $0.35897...$ and $0.89744...$, respectively.
For the HFP without reflexion we obtain $x_{F3}=0.74421$ compared to the exact value $0.74359$.
An analysis of the same time series with, for example, recurrence methods yields a relative accuracy
of $0.05$ for the position of the fixed points and is therefore much less accurate.
Our algorithm for the turning point analysis of time series yields reliable results even if the number
of points of the time series is comparatively small.

The above algorithm can also be applied to the case of a scalar experimental 1D time
series, i.e. a chaotic signal, which originates from a higher dimensional dynamics.
It is therefore by no means restricted to a 1D time series arising from a 1D system.
The oscillating behaviour around the unstable fixed points described above represent a characteristic
feature of each component of a multidimensional signal. Indeed the chaotic dynamics
of a multidimensional system in the neighbourhood of a fixed point can be characterized
by the fact that the deflection or turning of the different trajectories approaching
the fixed point varies significantly
and therefore turnings can occur with respect to any direction of space.
As examples for two-dimensional fully chaotic systems we mention the Ikeda [Ikeda,1979] or Henon map [Henon,1976]
which exhibits oscillatory behaviour with small amplitudes around the position of the fixed point for both
components $x$ as well as $y$.

\section{Summary and Outlook}

We have investigated the ergodic and fully developed chaotic behaviour of SSH maps in some detail.
Our starting point was the observation that the chaotic trajectories suggest the existence of certain
nontrivial structures around the unstable fixed point. A closer look at the trajectories showed us that these structures
have their origin in the distribution of oscillations in the unit interval. A comparison of the densities for
the center of oscillations as well as their amplitudes with the corresponding quantities for a weighted
random map clearly revealed that the dynamics of the oscillations is highly
organized and far from being random. Characteristics of the amplitude of oscillations as a function of the
center of oscillations have been discussed. 

The central idea of our investigation is the introduction of a turning point analysis as a tool
for the characterization of the dynamics of ergodic and fully developed chaotic systems. 
On our way of a quantification of this idea we have derived a turning point criterion which gives
us a condition for the point $x(n-1)$ so that $x(n)$ becomes a turning point. 
With the help of this criterion we were able to understand the dynamics of turning points. Roughly 
speaking the trajectories possess two phases of motion: a stretching phase in the neighbourhood
of the unstable fixed point $0$ which consists apart from its important first point only of non turning points
and an oscillating stretching phase around the second unstable fixed point $x_F$ which consists exclusively
of turning points.

As an important tool we have subsequently introduced the turning point map (TPM) which
illustrates the mapping of successive turning points. This TPM exhibits a number of dynamically
interesting properties. It contains details of the first hump of iterates $f^{i}$ for all $i$ and maps in particular
arbitrarily small intervals in the neighbourhood of the
origin onto the finite interval $[x_F,1]$. The TPM obeys an asymptotic scaling law and converges to
a repeating functional form. A construction scheme for the humps in the asymptotic scaling regime has
been derived. 
Another important property of the TPM is the fact that it allows us to locate the fixed point $x_F$
as well as all its preimages. The turning point dynamics shows an
infinite correlation length of the corresponding correlation function, i.e. the turning point
analysis extracts the correlated part from the originally completely uncorrelated trajectory.
The origin of this strong correlation is the oscillating dynamics around the fixed point $x_F$.

We have shown that it is possible to define a so-called renormalized limit function which
corresponds to the renormalized version of the infinitely iterated map. The renormalization
is performed with respect to successive zeros. Successive humps are scaled onto the unit
interval. We have illustrated the construction scheme of the limit function for the example
of the logistic map.

As a next step we have investigated the TPMs belonging to higher iterates of the original map $f$. 
The TPMs themselves as well as their disconnected constituents show an asymptotic scaling behaviour
and exactly describe important parts of arbitrarily high iterates of the map $f$.
It is possible to extract the position of not too long periodic orbits of $f$ by considering the
TPM of the corresponding iterate $f^{n}$. A limited amount of information on the
stability of the periodic orbits can be obtained from the scaling laws of the constituents of the TPMs. 
Informations on the position of high preimages of the fixed points can easily be extracted.

We have performed a statistical analysis of the turning point dynamics for the special
case of doubly symmetric maps. We could show that
the ratio of turning points to the total number of points is an invariant for these
maps and takes on the value $\frac{2}{3}$. This suggests that the ratio might be used for a
characterization of nonsymmetric maps and divides the maps into different equivalence classes. 

Our investigation and discussion of the dynamical and statistical behaviour of the
turning points, the TPMs as well as the correlation functions clearly show that a turning point
analysis extracts valuable information from a given ergodic and fully chaotic system
and is very well-suited for the analysis of chaotic time series.
Using the main features of the dynamics of the turning points in the neighbourhood of the
fixed point we developed an algorithm which allows us to locate the unstable fixed points and higher
periodic orbits thereby distinguishing between hyperbolic fixed
points with and without reflection. The most important
property of our turning point method for the location of the fixed points is the fact that it is
not restricted to the linear neighbourhood of the fixed point. Even in the case when the linear
regime around the fixed point is not visited by the time series at all, it is possible to 
determine approximately the position of the fixed point. 
Our turning point analysis of the time series yields reliable results in particular if the number
of points of the time series is comparatively small and represents therefore an attractive method
for the investigation of dynamical systems of, for example, biological origin. 

\vspace*{1.0cm}

{\large \bf Acknowledgements}
 
The European Community (F.K.D.) is gratefully acknowledged for financial support.

\newpage

\begin{center}
{\large \bf References}
\end{center}

F.T.Arecchi, R.Meucci, G.Puccioni and J.Tredicce [1982],
"Experimental evidence of subharmonic bifurcations, multistability, and
turbulence in a Q-switched gas laser",\\
Phys.Rev.Lett.{\bf{49}}, 1217-1220\\

C.Beck and F.Schl\"ogl [1993], "Thermodynamics of chaotic systems", Cambridge
Nonlinear Science Series.\\

A.Blokh and M.Misiurewicz, "{\it New order for periodic orbits of
interval maps}", preprint\\

P.Collet and J.P.Eckmann [1980], "Iterated maps on the interval as dynamical
systems", Birkh\"auser, Cambridge Massachusetts\\

J.P.Crutchfield, J.D.Farmer and B.A.Huberman [1982],
"Fluctuations and simple chaotic dynamics", Phys.Rep.{\bf{92}}, 45-82\\

F.K.Diakonos and P.Schmelcher [1997], subm.to CHAOS\\

F.K.Diakonos and P.Schmelcher [1996],
"On the construction of one-dimensional iterative maps from the invariant
     density: the dynamical route to the beta distribution", Phys.Lett.A{\bf{211}}, 199 \\

F.K.Diakonos and P.Schmelcher, in preparation\\

E.J.Ding [1988], "Wave numbers for unimodal maps", Phys.Rev.A{\bf{37}}, 1827-1830\\

M.Feigenbaum [1978], "Quantitative universality for a class of nonlinear transformations",
J.Stat.Phys.{\bf{19}}, 25-52\\

M.Feigenbaum [1979], "The universal metric properties of nonlinear transformations",
J.Stat.Phys.{\bf{21}}, 669-706\\ 

J.P.Gollub and S.V.Benson [1980],
"Many routes to turbulent convection", J.Fluid Mech.{\bf{100}}, 449-470 \\

S.Grossmann and S.Thomae [1977],
"Invariant distributions and stationary correlation functions of
one-dimensional discrete processes", Zeitschrift f\"ur Naturforschung
{\bf{32}}, 1353-1363 \\

G.Gy\"orgyi and P.Szepfalusy [1984],
"Fully developed chaotic 1-d maps", Z.Phys.B{\bf{55}}, 179-186\\ 

G.Gy\"orgyi and P.Szepfalusy [1984],
"Properties of fully developed chaos in one-dimensional maps", J.Stat.Phys.{\bf{34}}, 451-475 \\

M.Henon [1976], "A Two-dimensional Mapping with a Strange Attractor", Comm.math. Phys.{\bf{50}}, 69-77\\

K.Ikeda [1979], "Multiple-valued stationary state and its instability of the 
transmitted light by a ring cavity system", Opt. Comm. {\bf 30}, 257

L.Kadanoff [1983], "Roads to chaos", Physics Today, p.46-53\\

P.S.Linsay [1981],
"Period doubling and chaotic behavior in a driven anharmonic oscillator", Phys.Rev.Lett.{\bf{47}}, 1349-1352 \\

E.Ott [1993], "Chaos in dynamical systems", Cambridge Univ. Press, Cambridge\\

S.Pincherle [1920], "On the complete iteration of $x^2-2$", Rend. della Real Acad. dei Lincei {\bf{29}},
329-333\\

R.Z.Sagdeev, D.A.Usikov and G.M.Zaslavsky [1990], "Nonlinear Physics", Harwood
Academic Publishers, New York\\

H.G.Schuster [1994], "Deterministic Chaos", VCH Weinheim\\

\newpage


\begin{figure}[ht]
\centerline{\psfig{figure=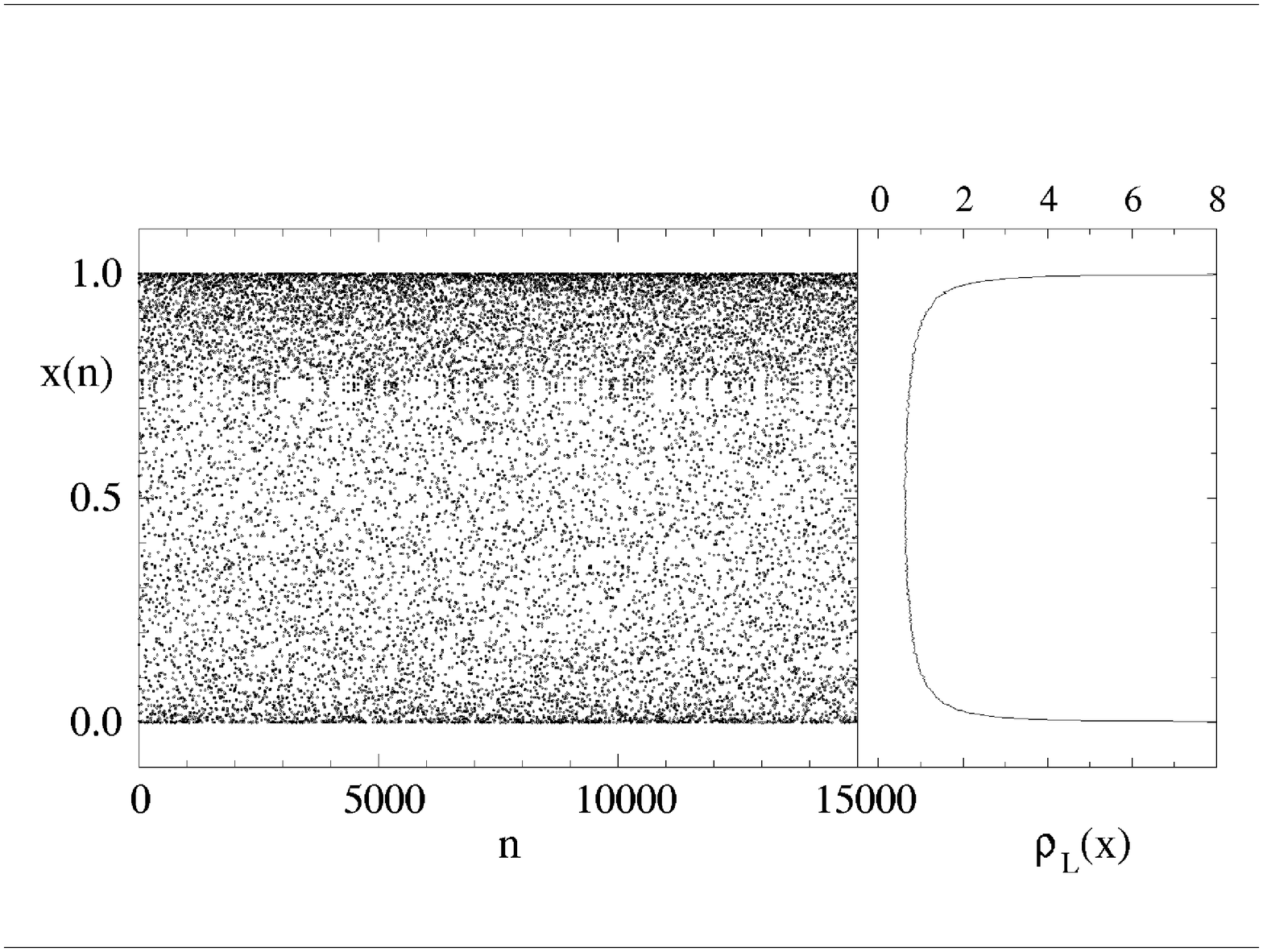,width=\textwidth,angle=270}}

\vspace{1.0cm}

{\bf{Figure 1:}}
A chaotic and ergodic trajectory $x(n)$ of the logistic map for $1.5 \times 10^{4}$ iterations
and the invariant density $\rho_{L}$ of the same trajectory for $10^{6}$ iterations.
\end{figure}

\begin{figure}[ht]
\centerline{\psfig{figure=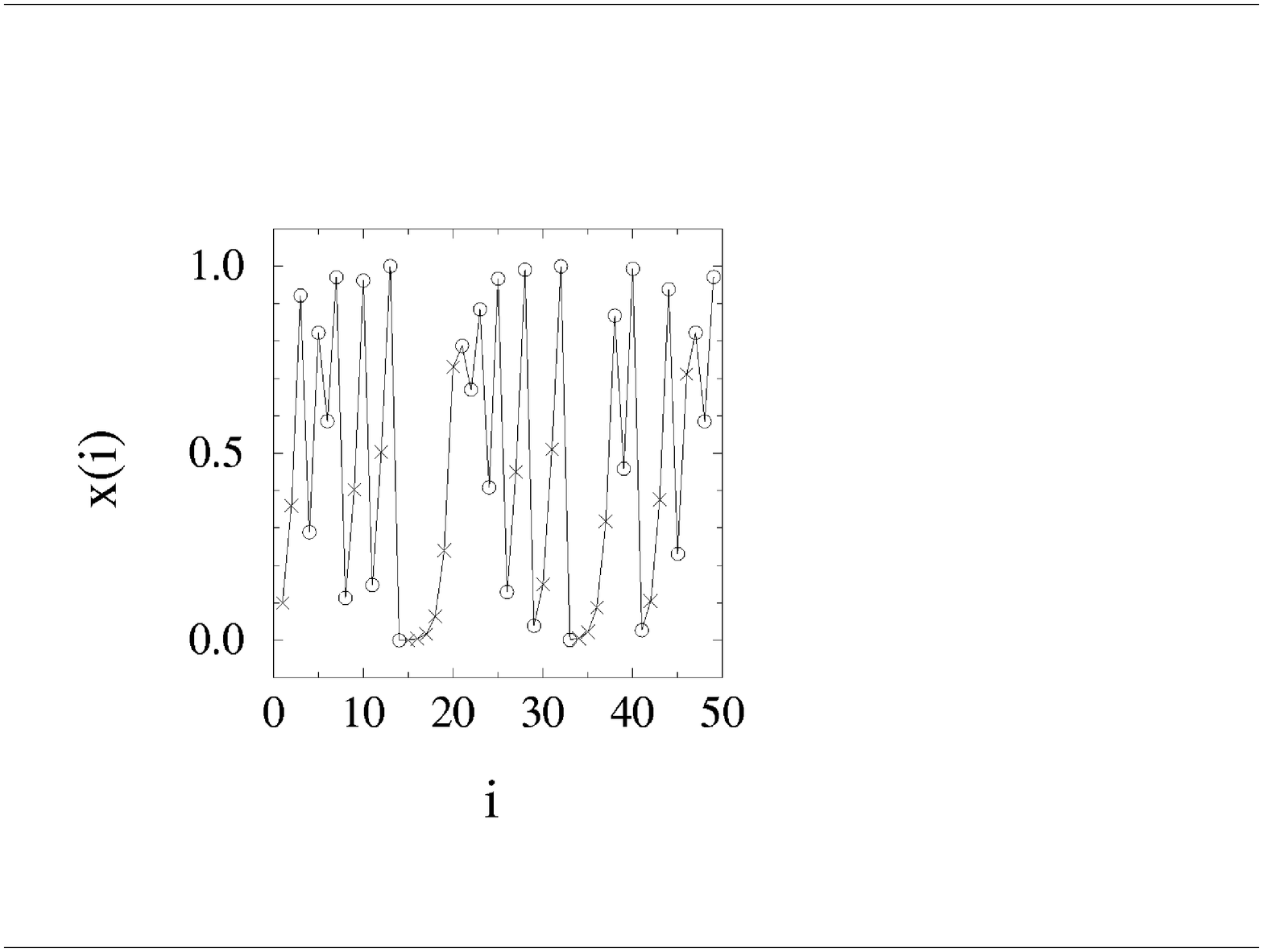,width=\textwidth,angle=270}}

\vspace{1.0cm}

{\bf{Figure 2:}}
50 iterations of a trajectory of the logistic map. The turning points are indicated with circles whereas
the non-turning points are indicated with crosses. The stretching and oscillatory phases of motion are
clearly visible.

\end{figure}

\begin{figure}[ht]
\centerline{\psfig{figure=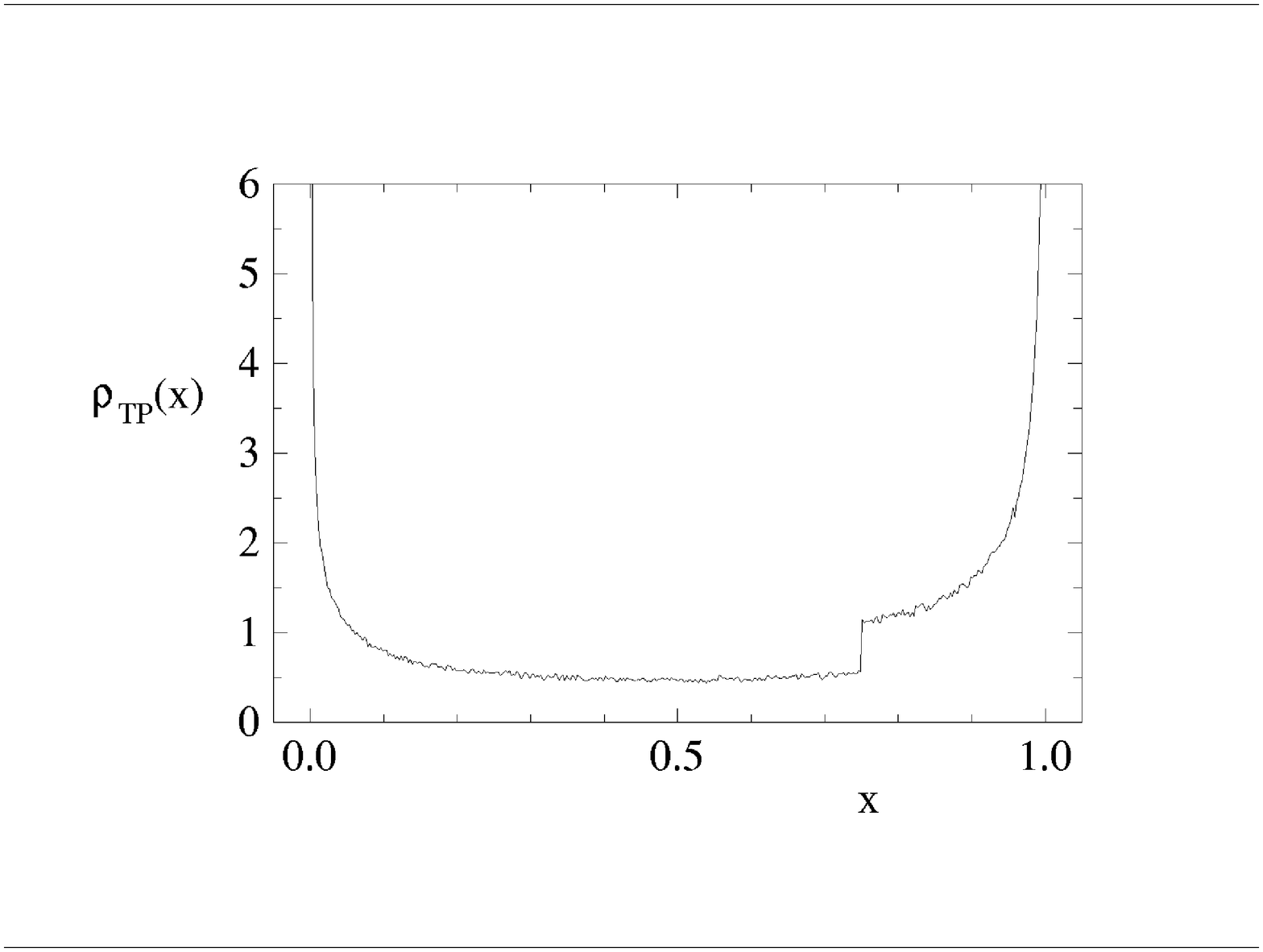,width=\textwidth,angle=270}}
\vspace{1.0cm}

{\bf{Figure 3:}}
The turning point density $\rho_{TP}$ of the logistic map. The step like structure is at the position of the
fixed point $x_F$. 

\end{figure}

\begin{figure}[ht]
\centerline{\psfig{figure=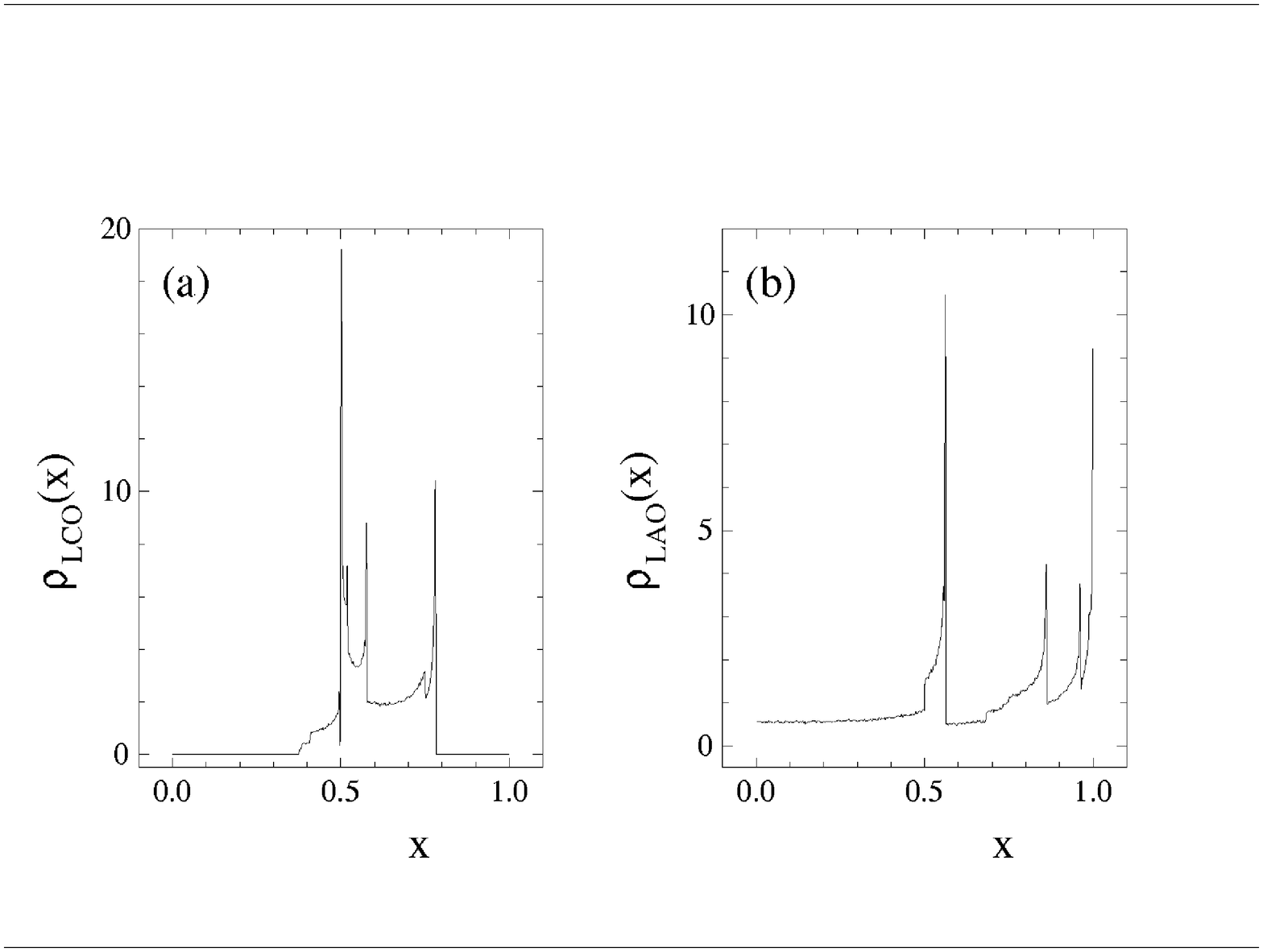,width=\textwidth,angle=270}}
\vspace{1.0cm}

{\bf{Figure 4:}}
(a) The density for the centers of the oscillations $\rho_{LCO}(x)$ of a typical chaotic trajectory of the 
logistic map.
(b) The density for the amplitudes of the oscillations $\rho_{LAO}(x)$ of the same trajectory. 

\end{figure}

\begin{figure}[ht]
\centerline{\psfig{figure=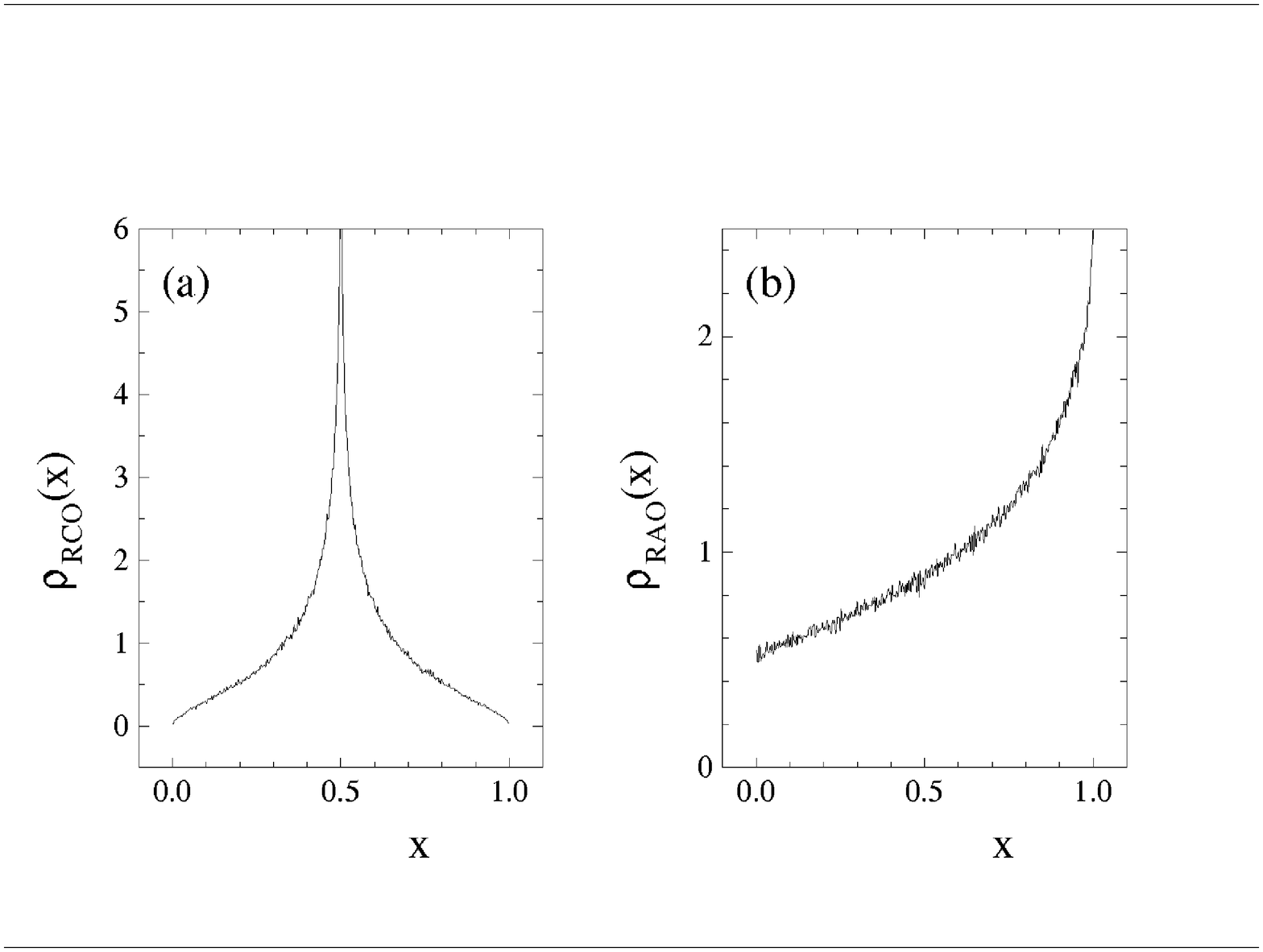,width=\textwidth,angle=270}}
\vspace{1.0cm}

{\bf{Figure 5:}}
(a) The density for the centers of the oscillations $\rho_{RCO}(x)$ of a trajectory of the 
weighted random map with $\rho_R = \rho_L$.
(b) The density for the amplitudes of the oscillations $\rho_{RAO}(x)$ of the same trajectory. 

\end{figure}

\begin{figure}[ht]
\centerline{\psfig{figure=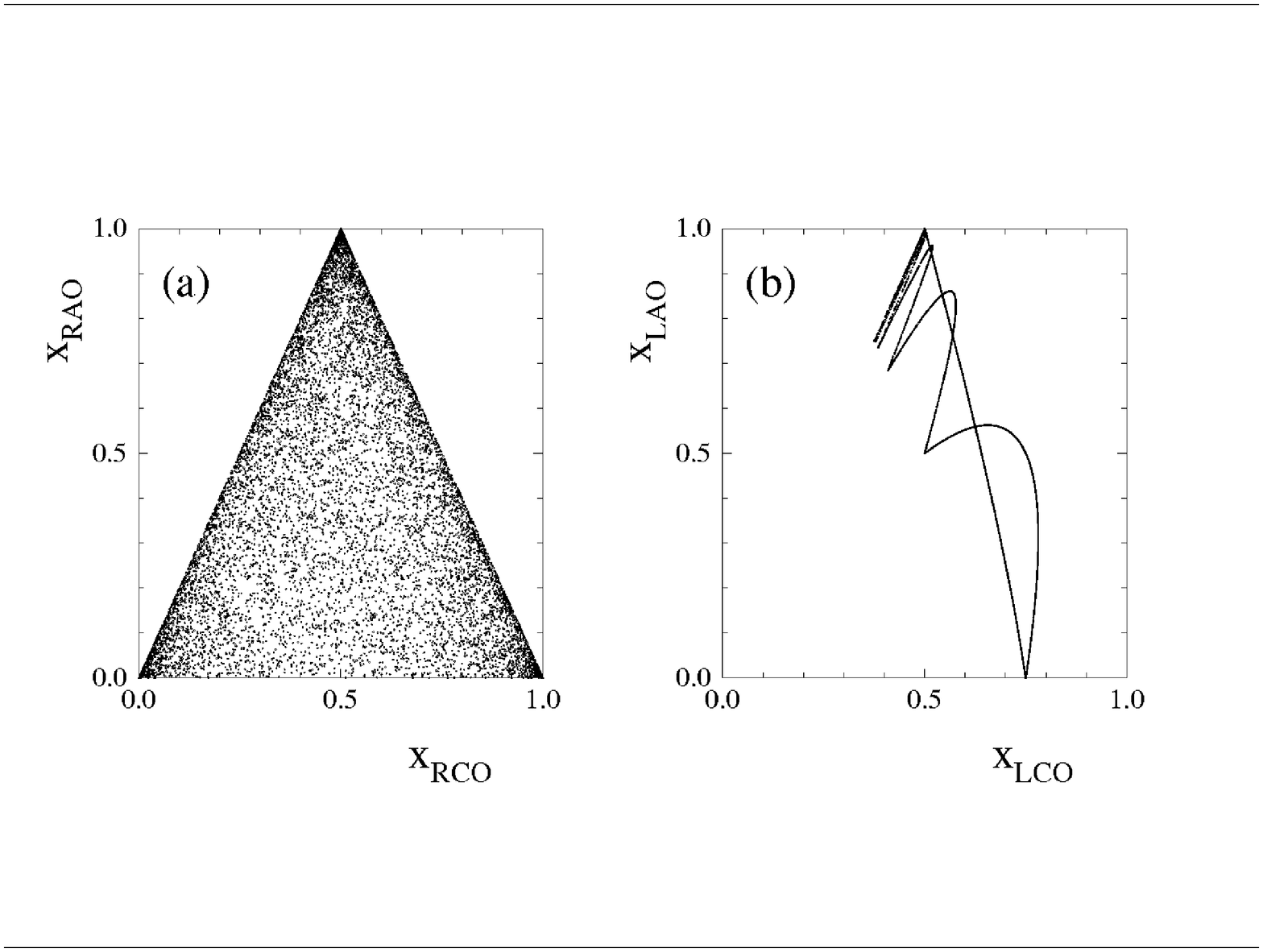,width=\textwidth,angle=270}}
\vspace{1.0cm}

{\bf{Figure 6:}}
(a) The amplitudes of the oscillations as a function of the centers of the oscillations $x_{RAO}(x_{RCO})$
for the random map.
(b) The amplitudes of the oscillations as a function of the centers of the oscillations $x_{LAO}(x_{LCO})$
for the logistic map.

\end{figure}

\begin{figure}[ht]
\centerline{\psfig{figure=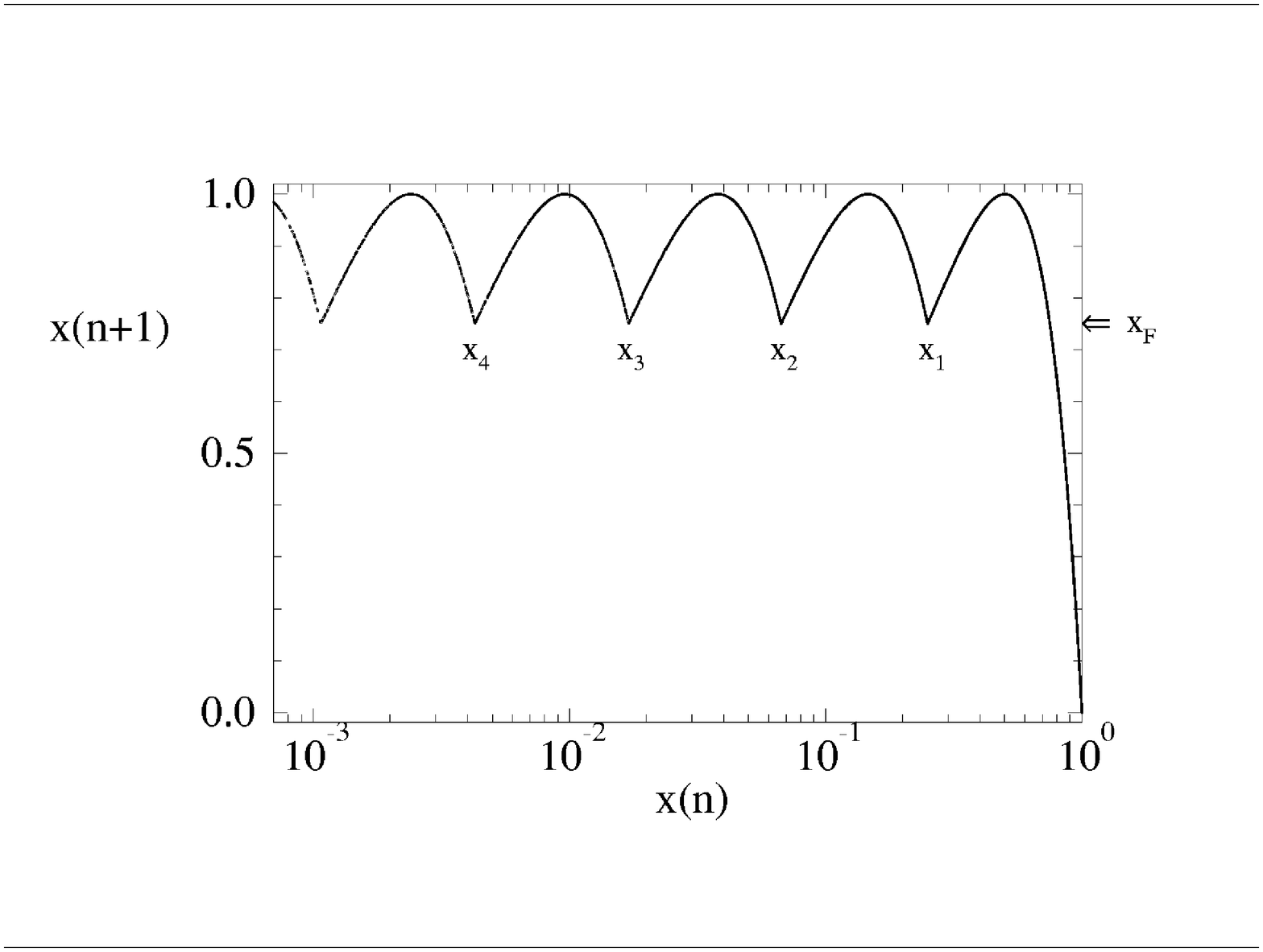,width=\textwidth,angle=270}}
\vspace{1.0cm}

{\bf{Figure 7:}}
The turning point map of the logistic map for a logarithmic abscissae.
$x_i$ are successive preimages of the unstable
fixed point $x_F=\frac{3}{4}$ on the left branch of $f$. 

\end{figure}

\begin{figure}[ht]
\centerline{\psfig{figure=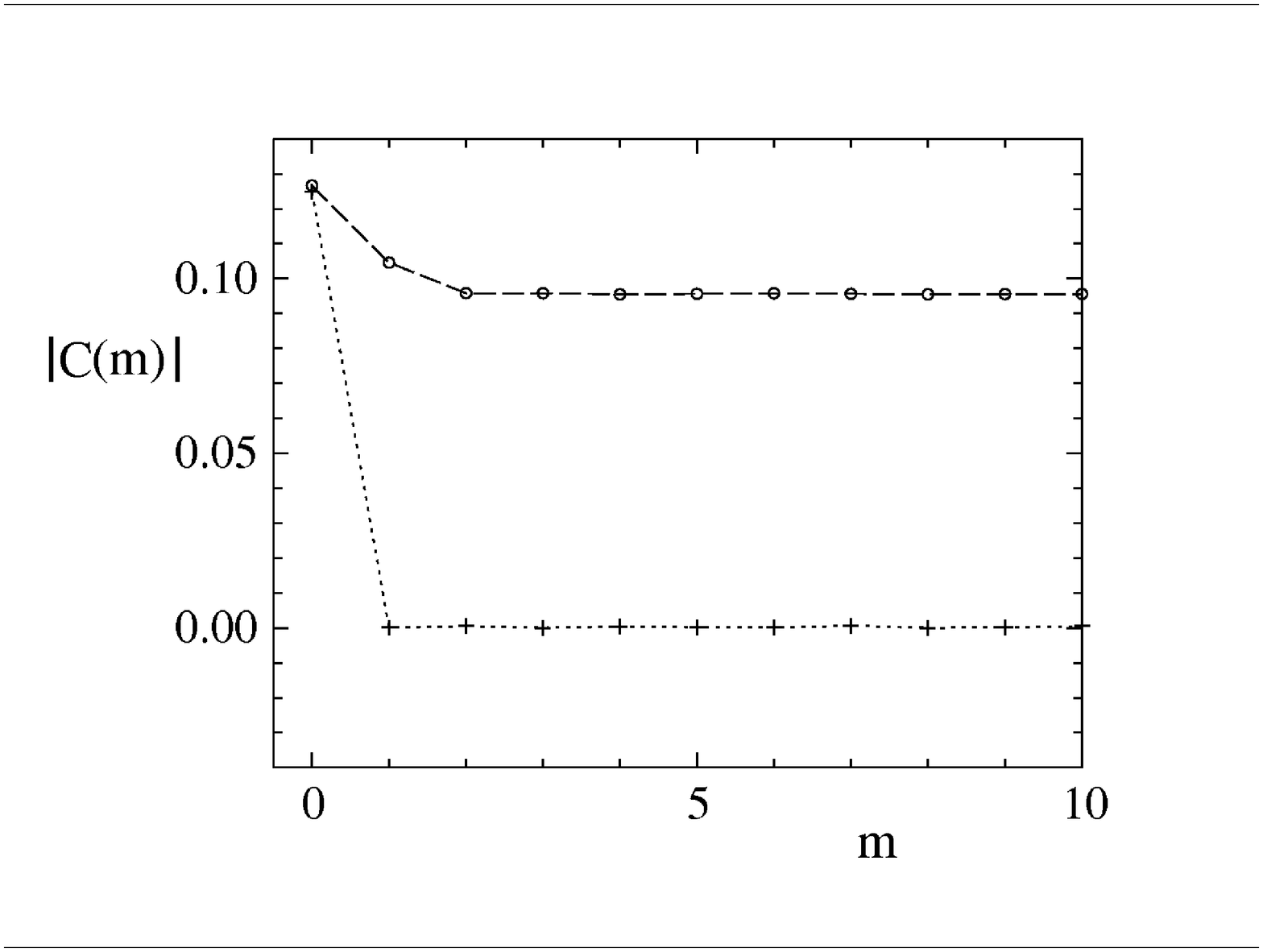,width=\textwidth,angle=270}}
\vspace{1.0cm}

{\bf{Figure 8:}}
The absolute value $|C(m)|$ of the correlation function 
$C(m) = \langle (x-\langle x \rangle_{f})(f^{m}(x)-\langle x \rangle_{f})\rangle_{f}$
for the logistic map. Circles indicate the correlation function of the turning points
and crosses the correlation function of the original trajectory.

\end{figure}

\begin{figure}[ht]
\centerline{\psfig{figure=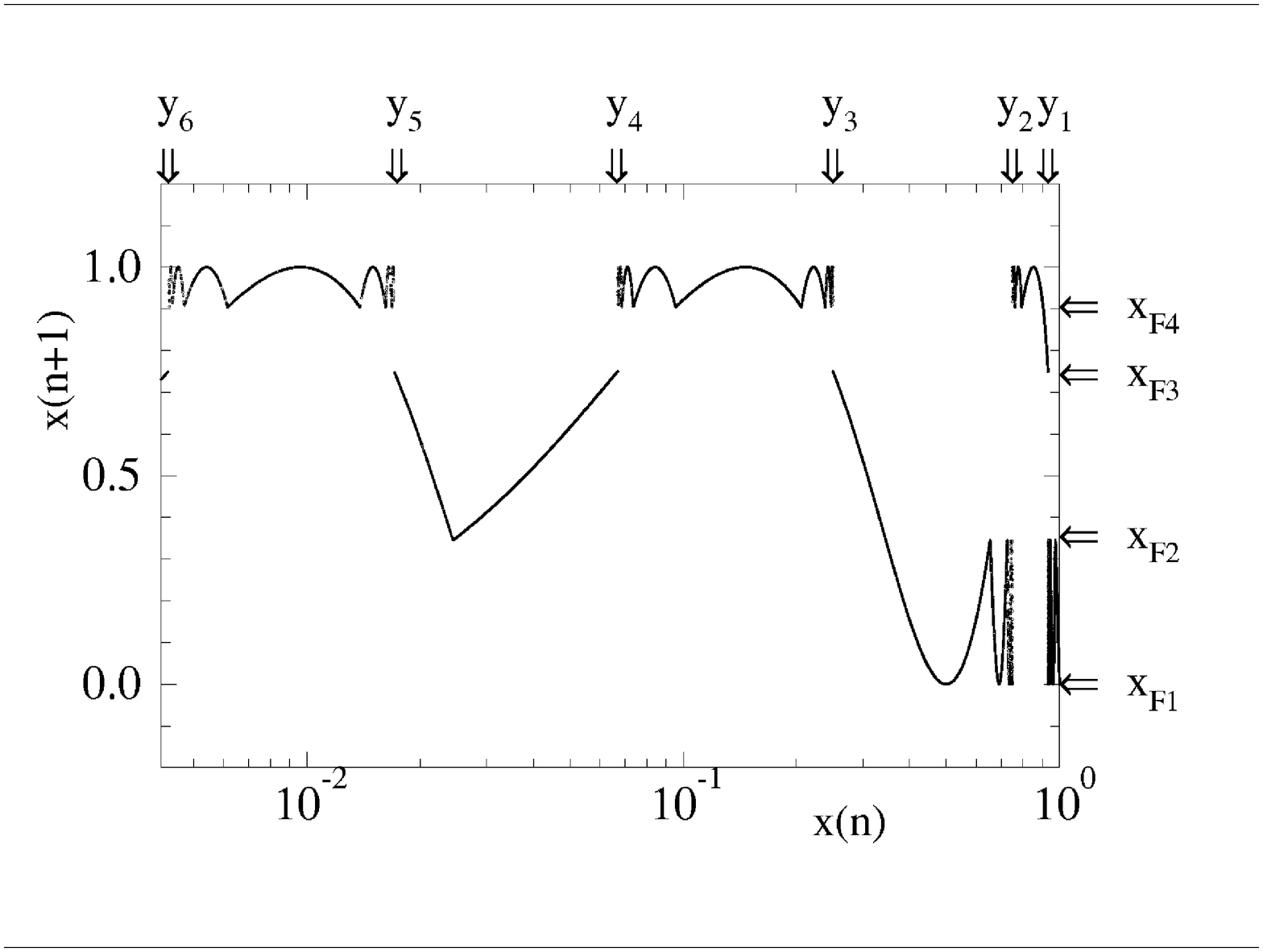,width=\textwidth,angle=270}}
\vspace{1.0cm}

{\bf{Figure 9:}}
The turning point map of the doubly iterated logistic map $f^{2}$ for a logarithmic abscissae.
The disconnected objects
$(a),(e)$ repeat themselves in an asymptotically scaling way if we approach the unstable
fixed point $0$. All objects show also an interior scaling behaviour (see Figure 9).
$y_i$ correspond to the positions of the preimages of the fixed point with a positive derivative,
i.e. preimages of $x_{F3}=\frac{3}{4}$. The positions of the fixed points $x_{Fi}$ of $f^{2}$ are
indicated on the vertical axis.   

\end{figure}

\begin{figure}[ht]
\centerline{\psfig{figure=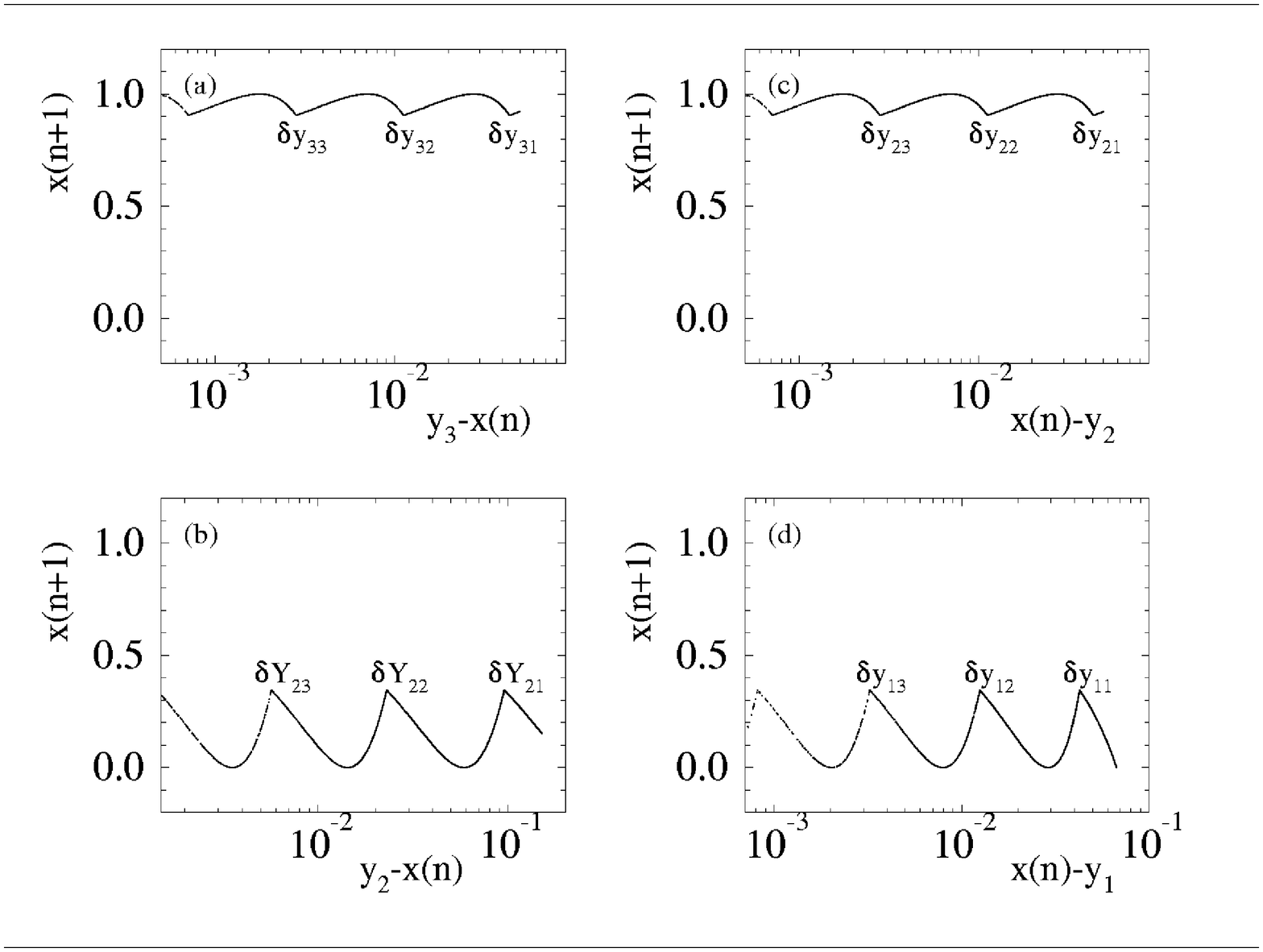,width=\textwidth,angle=270}}
\vspace{1.0cm}

{\bf{Figure 10:}}
Magnified parts of the objects $(a),(b),(c),(d)$ of Figure 8 are shown on a logarithmic scale.
The scaling behaviour of the interior of these objects becomes evident.
(a) Scaling behaviour towards the preimage $y_3$ of $x_{F3}$ by a series $\delta y_{31},\delta y_{32},\delta y_{33},...$
of preimages of the fixed point $x_{F4}$.
(b) Scaling behaviour towards the preimage $y_2$ of $x_{F3}$ by a series $\delta Y_{21},\delta Y_{22},\delta Y_{23},...$
of preimages of the fixed point $x_{F2}$.
(c) Scaling behaviour towards the preimage $y_2$ of $x_{F3}$ by a series $\delta y_{21},\delta y_{22},\delta y_{23},...$
of preimages of the fixed point $x_{F4}$.
(d) Scaling behaviour towards the preimage $y_1$ of $x_{F3}$ by a series $\delta y_{11},\delta y_{12},\delta y_{13},...$
of preimages of the fixed point $x_{F2}$.

\end{figure}

\begin{figure}[ht]
\centerline{\psfig{figure=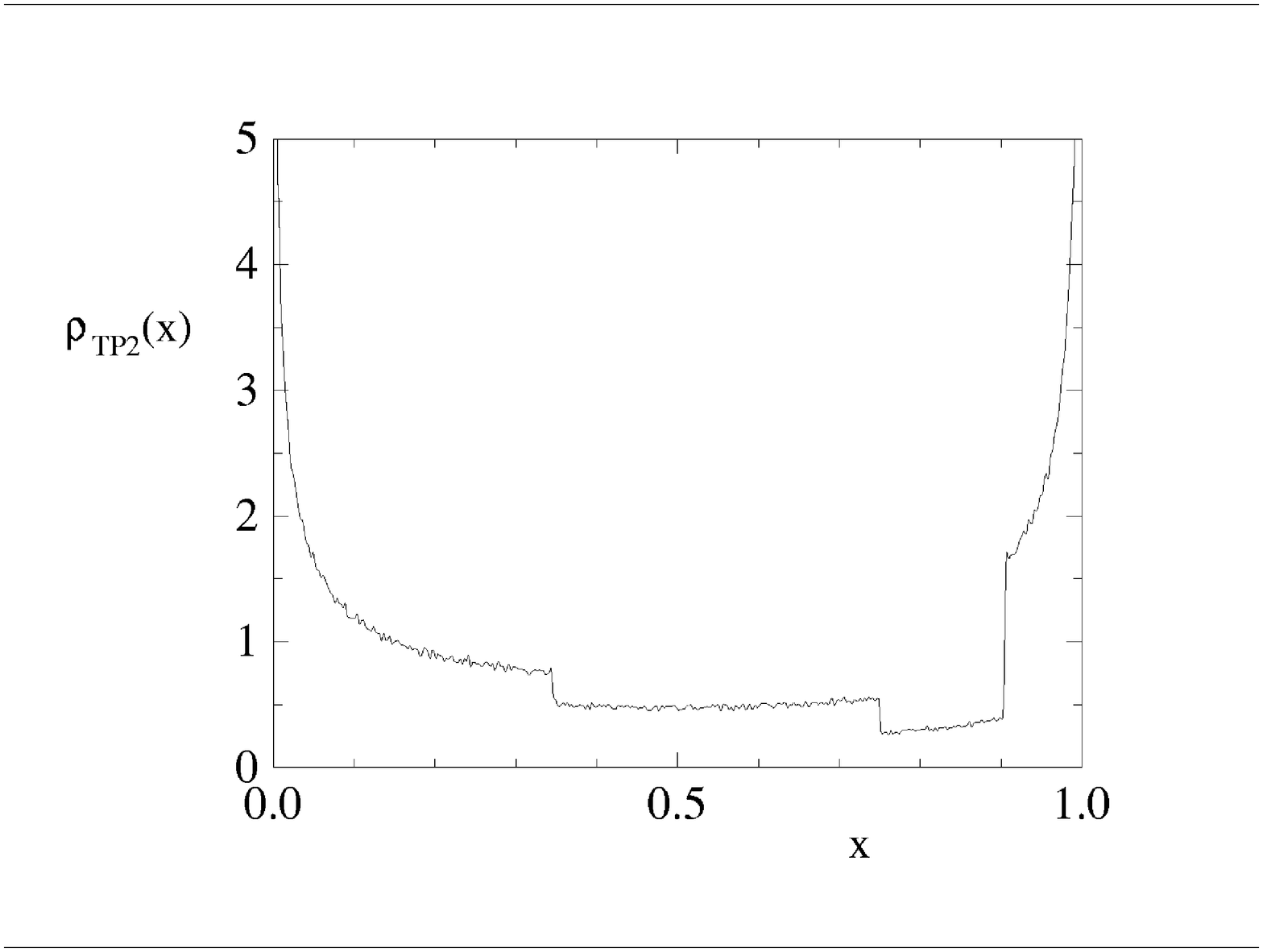,width=\textwidth,angle=270}}
\vspace{1.0cm}

{\bf{Figure 11:}}
The density $\rho_{TP2}$ of the turning points of the doubly iterated logistic map $f^{2}$.
Three step like structure are clearly visible. The positions of these steps are the 
fixed points $x_{Fi}$ of $f^{2}$.

\end{figure}

\end{document}